\documentclass[preprint,aps,amsmath,
nofootinbib,floatfix]{revtex4}
\usepackage{subfigure}
\usepackage{graphicx}
\usepackage{epstopdf}
\usepackage{multirow}

\usepackage{slashed}

\usepackage{color}
\usepackage{comment}
\usepackage{booktabs} % This is for the Table Line Width \toprule[1pt]
\usepackage{natbib}
\usepackage[colorlinks,linkcolor=black,anchorcolor=blue,citecolor=green]{hyperref}
\makeatletter

\newcommand{\Rmnum}[1]{\expandafter\@slowromancap\romannumeral #1@}
\makeatother
\newcommand{\met}{\ensuremath{{\not\mathrel{E}}_T}}
\newcommand{\gev}{{\rm GeV}}

 \def\sL{\ensuremath{\tilde t_L}}
\def\sR{\ensuremath{\tilde t_R}}
 \def\bL{\ensuremath{\tilde b}_L}
  \def\bR{\ensuremath{\tilde b}_R}
\def\met{\ensuremath{\slashed{E}_T}}
\def\n{\ensuremath{\chi^0_1}}
\def\nn{\ensuremath{\chi^0_2}}

\def\cha{\ensuremath{\chi^+_1}}

\newcommand{\minigraph}[5][0.25in]{\begin{minipage}{#2}\begin{center}\includegraphics[width=#2]{#5}\\\vspace{#3}\hspace{#1}{\footnotesize #4}\end{center}\end{minipage}}

\begin{document}
\begin{flushright}
PITT-PACC-1506
\end{flushright}

\preprint{APS/123-QED}
%\listoftables
\date{\today}

\title{Sbottom discovery via mixed decays at the LHC}
\author{
        Tao Han$^{1,2}$\footnote{than@pitt.edu},
        Shufang Su$^{3,2}$\footnote{shufang@email.arizona.edu},
        Yongcheng Wu$^{2}$\footnote{wuyongcheng12@mails.tsinghua.edu.cn  },
         Bin Zhang$^{2}$\footnote{zb@mail.tsinghua.edu.cn},
         Huanian Zhang$^{3}$\footnote{fantasyzhn@email.arizona.edu}
         }

\affiliation{
$^{1}$ PITT PACC, Department of Physics and Astronomy, University of Pittsburgh,  Pittsburgh, PA 15260, USA\\
$^{2}$ Department of Physics, Tsinghua University, Beijing,  100086, China\\
$^{3}$ Department of Physics, University of Arizona, Tucson, Arizona  85721, USA
}

\begin{abstract}
In the search for bottom squark (sbottom) in SUSY at the LHC, the common practice has been to assume a $100\%$ decay branching fraction for a given search channel. 
In realistic MSSM scenarios, there are often more than one significant decay modes to be present, 
which significantly weaken the current sbottom search limits at the LHC. On the other hand,  the combination of the multiple decay modes offers alternative discovery channels for sbottom searches.  In this paper, we present the sbottom decays in a few representative mass parameter scenarios.  We then analyze the sbottom signal for the pair production in QCD with one sbottom decaying via $\tilde{b}\rightarrow b \chi_1^0,\  b \chi_2^0$, and the other one decaying via $\tilde{b} \rightarrow t \chi_1^\pm$.  With the gaugino subsequent decaying to gauge bosons or a Higgs boson $\chi_2^0 \rightarrow Z \chi_1^0,\ h \chi_1^0$ and $\chi_1^\pm \rightarrow W^\pm \chi_1^0$, we study the reach of those signals at the 14 TeV LHC with 300 ${\rm fb^{-1}}$ integrated luminosity.
For a left-handed bottom squark, we find that a mass up to 920 GeV can be discovered at 5$\sigma$ significance for 250 GeV $< m_{\chi_1^0} <$ 350 GeV, or excluded up to 1050 GeV at the 95\% confidence level for the $h$ channel ($\mu>0$); similarly, it can be discovered  up to 840 GeV, or excluded up to 900 GeV at the 95\% confidence level for the $Z$ channel ($\mu<0$). The top squark reach is close to that of the bottom squark. The sbottom and stop signals in the same SUSY parameter scenario are combined to obtain the optimal sensitivity, which is about 150 GeV better than the individual reach of the sbottom or stop. 
For a right-handed bottom squark with  $\tilde{b} \tilde{b}^* \rightarrow b \chi_1^0,\  t \chi_1^\pm$ 
channel, we find that the sbottom mass up to 880 GeV can be discovered at 5$\sigma$ significance, or excluded up to 1060 GeV at the 95\% confidence level.

\end{abstract}

\maketitle

\section{Introduction}

The outstanding performance of the Large Hadron Collider (LHC) at CERN had led to the milestone discovery of the Higgs boson predicted by the Standard Model (SM). High energy physics has thus entered a new era in understanding the nature of electroweak symmetry breaking. ``Naturalness'' argument for the Higgs boson mass implies new physics associated with the SM Higgs sector not far above the TeV scale \cite{Feng:2013pwa,Giudice:2013yca,Altarelli:2013lla}.
%Admittedly, it has been disturbing that there have not been clear indications for the existence of new physics beyond the Standard Model. 
The LHC Run-2 with higher energy and higher luminosity will certainly extend the horizon to seek for new physics. Among the new physics scenarios, the weak scale Supersymmetry (SUSY) remains to be the most attractive option because of the accommodation for a light Higgs boson, the natural dark matter candidate, and the possibility for gauge coupling unification. In preparing to exploit a large amount of the incoming data from the LHC experiments, it is thus of priority to embrace the SUSY searches in a comprehensive way. 

While the top squark (stop $\tilde{t}$) sector might be the most relevant supersymmetric partner in connection to the Higgs physics given the large top Yukawa coupling, the bottom squark (sbottom $\tilde{b}$) 
sector is also of great interest.  The left-handed sbottom mass is related to the left-handed stop mass since they are controlled by the same soft SUSY breaking mass parameter \cite{Martin:1997ns,Chung:2003fi}. In the region of large ratio of the Higgs vacuum expectation values: $\tan\beta=v_2/v_1$, the bottom Yukawa coupling is large and there could be  large corrections to the Higgs physics from the sbottom sector as well \cite{Carena:2005ek}.    
Although the LHC program has been carrying out a rather broad and impressive SUSY search plan, 
%n the other hand, 
many searches are still under strong assumptions for the sake of simplicity.
The current sbottom search mainly focuses on the direct decay to the lightest supersymmetric particle (LSP) $\tilde{b} \rightarrow b \chi_1^0$, with $b \bar b+\met$ being the dominant search channel. % \cite{AdeelAjaib:2011ec}. 
With the data collected at the LHC 7 and 8 TeV, a sbottom with mass up to 700 GeV has been excluded in this channel \cite{CMS-PAS-SUS-13-018,Aad:2013ija,Aad:2014nra}. Even in the parameter space with highly degenerate sbottom and LSP masses \cite{Alvarez:2012wf,Lee:2012sy,Bi:2012jv,Dutta:2015hra,AdeelAjaib:2011ec}, a sbottom is excluded with mass up to about 255 GeV \cite{Aad:2014nra}.
%AdeelAjaib:2011ec. 
%\Yongcheng{It seems only this one for sbottom, there are several others for stop.}
%
Other decay channels including the cascade decay via the next-lightest supersymmetric particle (NLSP)  $\tilde{b} \to b \nn \rightarrow bh \n,\ bZ\n $~\cite{Aad:2014lra,CMS-PAS-SUS-13-008}  and  $\tilde{b} \to t W \n$~\cite{Aad:2014pda,Chatrchyan:2013fea,CMS-PAS-SUS-13-008,Chatrchyan:2014aea} have also been considered with a $100\%$ branching fraction each, with considerably weaker limits. 

%In the search for bottom squark (sbottom) in SUSY at the LHC, the common practice has been to assume a $100\%$ decay branding fraction for a given search channel. 

In realistic MSSM scenarios, there are often more than one significant decay modes to be present. Two prominent examples stand out: A left-handed sbottom in the Wino-NLSP scenario may have a decay $\tilde{b} \to b \nn$ with branching fraction as high as $30\% - 40\%$, along with the leading decay $\tilde{b} \rightarrow t \chi_1^\pm$; Similarly, a right-handed sbottom in the Higgsino-NLSP scenario may have the leading decay mode $\tilde{b} \to b \n$ with branching fraction only $40\% - 60\%$, along with a sub-leading decay $\tilde{b} \rightarrow t \chi_1^\pm$ of $20\% - 30\%$.
Those additional channels dilute the leading signals currently being searched for at the LHC, and 
significantly weaken the sbottom search limits when assuming $100\%$ branching fraction for a given search channel. On the other hand, the combination of the multiple decay modes offers alternative discovery channels for sbottom searches, that must be properly  taken into account.  
%\Tao{Some early work along this line exists in the literature \cite{Lee:2012sy}. (cannot miss literature.)}

In this paper, we present the sbottom decays in a few representative SUSY mass scenarios. We then analyze the sbottom pair production signal with one sbottom decaying via $\tilde{b}\rightarrow b \chi_1^0,\  b \chi_2^0$, and the other one decaying via $\tilde{b} \rightarrow t \chi_1^\pm$.  With the subsequent decay of $\chi_2^0 \rightarrow Z \chi_1^0,\ h \chi_1^0$ and $\chi_1^\pm \rightarrow W^\pm \chi_1^0$, we study the reach of those signals at the 14 TeV LHC with 300 ${\rm fb^{-1}}$ integrated luminosity.
% of $bbbbjj\ell+\met$ and $bbjjjj\ell\ell+\met$ final states.  
%We note that the left-handed squark parameter is common for the stop and sbottom
Because of the similarity of the final state signatures and potential correlation of the left-handed soft mass, the sbottom and stop signals are combined to obtain the optimal sensitivity for the  same SUSY parameter region. 
We find for a left-handed bottom squark, a mass up to 920 GeV can be discovered at 5$\sigma$ significance for 250 GeV $< m_{\chi_1^0} <$ 350 GeV, or excluded up to 1050 GeV at the 95\% confidence level for the $h$ channel ($\mu>0$); similarly, the bottom squark can be discovered  up to 840 GeV, or excluded up to 900 GeV at the 95\% confidence level for the $Z$ channel ($\mu<0$), the top squark reach is close to that of the bottom squark. The sbottom and stop signals in the same SUSY parameter scenario are combined to obtain the optimal sensitivity, which is about 150 GeV better than the individual reach of the sbottom or stop. 
For a right-handed bottom squark with the channel $\tilde{b} \tilde{b}^* \rightarrow b \chi_1^0,\  t \chi_1^\pm \to tbW+\met$, we find that a mass up to 880 GeV can be discovered at 5$\sigma$ significance, or excluded up to 1060 GeV at the 95\% confidence level.

The rest of the paper is organized as follows. In Sec.~\ref{sec:MSSM_sbottom}, we briefly present the sbottom sector in the MSSM and introduce the mass and mixing parameters. We then calculate the sbottom decays for various neutralino/chargino mass spectra. Assuming one decay channel dominant at a time, we summarize the current LHC stop and sbottom search results from both ATLAS and CMS experiments. In Sec.~\ref{sec:analyses}, we investigate the reach of the sbottom signal with mixed decay channels at the 14 TeV LHC. We combine the left-handed sbottom and stop signals for the same SUSY parameter region. 
%via final states with a Higgs or a $Z$.  
In Sec.~\ref{sec:conclusion}, we summarize our results. 

\section{MSSM sbottom sector}
\label{sec:MSSM_sbottom}

We work in the MSSM and focus primarily on the third generation squark sector.  We decouple other SUSY particles: the gluino, sleptons, and the first two generations of squarks. We also decouple the non-SM Higgs particles by setting $M_A$ large.  Besides the third generation squarks, the other relevant SUSY states are a Bino (with a soft SUSY breaking mass $M_1$),  Winos (with a soft SUSY breaking mass $M_2$), and Higgsinos (with bilinear Higgs mass parameter $\mu$). Up on the mass diagonalization, they form neutralinos ($\chi_{1,2,3,4}^0$) and charginos ($\chi_{1,2}^\pm$).

\subsection{The sbottom sector}
\label{sec:sb}

The gauge eigenstates for the the third generation squark sector are $\tilde{t}_L, \tilde{b}_L, \tilde{t}_R, \tilde{b}_R$, where the left-handed states form a SU(2)$_L$ doublet with the soft SUSY breaking mass $M_{3SQ}$, and the right-handed states are SU(2)$_L$ singlets with soft SUSY breaking masses $M_{3SU},\ M_{3SD}$. For the sbottom sector, the mass matrix in the basis of  $(\tilde{b}_L,  \tilde{b}_R)$ is \cite{Martin:1997ns,Chung:2003fi}
\begin{equation}
M_{\tilde{b}}^2 = \begin{pmatrix}
M_{3SQ}^2+m_b^2+\Delta_{\tilde{d}_L} & m_b\tilde{A}_b \\
m_b\tilde{A}_b & M_{3SD}^2+m_b^2+\Delta_{\tilde{d}_R} \\
\end{pmatrix},
\end{equation}
where 
\begin{equation}
\begin{split}
\Delta_{\tilde{d}_L} = (-\frac{1}{2}+\frac{1}{3}\sin^2\theta_W)\cos2\beta M_Z^2, \quad
\Delta_{\tilde{d}_R} = \frac{1}{3}\sin^2\theta_W\cos2\beta M_Z^2
\end{split}
\end{equation}
are the contributions from the SU(2)$_L$ and U(1)$_Y$ D-term quartic interactions. The trilinear soft SUSY breaking coupling $A_b$ leads to the off-diagonal term $\tilde{A}_b=A_b-\mu\tan\beta$, that induces the mixing between left-handed and right-handed sbottom states. The lighter and heavier mass eigenvalues will be denoted as $m_{\tilde{b}_1},\ m_{\tilde{b}_2}$, respectively. 

The left-handed mass parameter $M_{3SQ}$ also controls the mass of the lighter stop.  
Since the stop sector provides the dominant contribution to the Higgs mass corrections, ``naturalness'' argument prefers a relatively lower value of the stop mass. 
%we decouple the  right-handed sbottom in our analysis.  
It is thus reasonable to consider $m_b \tilde A_b, M_{3SQ}^2 < M_{3SD}^2$, 
and the lighter sbottom mass eigenstate is mostly left-handed $\tilde{b}_1 \sim \bL$.  
The left-handed sbottom couples to a bottom quark and a neutralino (or a top quark and a chargino) mainly through SU(2)$_L$ gauge coupling and top Yukawa coupling, depending on the components of the neutralinos/charginos.   
Although the sbottom corrections to the Higgs mass are small compared to that from the stop, there can be significant modification to the Higgs couplings, especially  the bottom Yukawa coupling~\cite{Carena:2005ek,Belyaev:2013rza}. 

A right-handed sbottom $\bR$, on the other hand, couples to the U(1)$_Y$ gaugino and Higgsinos only via the hyper-charge and Yukawa coupling. Its mass is determined by $M_{3SD}^2$. We will also consider the situation when it is light.

\subsection{Sbottom decays}
\label{sec:decay}

The most commonly studied channel in experimental searches is the case $\tilde{b}_1 \rightarrow b \n$ with a branching fraction of $100\%$. This is true for the case with the Bino-LSP and the sbottom-NLSP, or the case with the stop-NLSP but $m_{\tilde{b}} <  m_{\tilde{t}} +M_W$, 
or the case with the Wino-NLSP for a right-handed sbottom. 
In a more general ground, sbottom decays lead to a much richer pattern.
%being the only electroweakino that is lighter than the sbottom, and minimal mixing in the stop sector, 

\subsubsection{The decay of $\bL$}

For a more general electroweakino spectrum, other decay channels may appear or even dominate, as analyzed in detail in Ref.~\cite{Eckel:2014wta}. 
We first consider the case of the lighter sbottom being mainly left-handed $\tilde{b}_1 \sim \bL$.
The mass spectrum of sbottom and gaugino would influence severely the decay modes of sbottom. We discuss the  (mainly left-handed) sbottom decay in details in the two general situations with a Bino-LSP:
\begin{eqnarray}
&& m_{\tilde{b}_1} > M_2 >M_1 \quad ({\rm Wino-NLSP}), \\
&& m_{\tilde{b}_1} > |\mu|>M_1 \quad ({\rm Higgsino-NLSP}). 
\end{eqnarray}
% $m_{\tilde{b}} > M_2>|\mu|>M_1$? BR versus msb is one option, just like those shown in E-S-Z's stop paper. Could also show BR versus (a) $M_2$ and (b) $\mu$ for a fixed msg (650 GeV?)})   When there is a large left-right mixing in the stop sector, $\tilde{t}_1$ could be lighter than $\tilde{b}_1$ such that $\tilde{b}_1 \rightarrow W \tilde{t}_1$ open and could dominate.  The cascade decay of the $\tilde{t}_1$ leads to complicated final states  for the $\tilde{b}_1$ signal as well.  
The more involved cases when both Winos and Higgsinos are below the sbottom mass threshold 
\begin{eqnarray}
&& m_{\tilde{b}_1} > |\mu|>M_2>M_1 \quad ( {\rm Wino-NLSP/Higgsino-NNLSP}), \\
&& m_{\tilde{b}_1} > M_2 > |\mu|>M_1 \quad ( {\rm Higgsino-NLSP/Wino-NNLSP}),
\end{eqnarray}
are also included when distinct features are present (sometimes referred as mixed NLSP's).  
% \Shufang{The name is a bit confusing.  For Eq(5), it is still Wino NLSP, only that Higgsino is also lighter than sbottom.  It is a bit confusing to call it Wino-Higgsino-NLSP.  But I guess I can live with it if there is no better names.}
 
\begin{figure}[tb]
\minigraph{12.5cm}{-0.3in}{(a)\hspace{2 in}(b)}{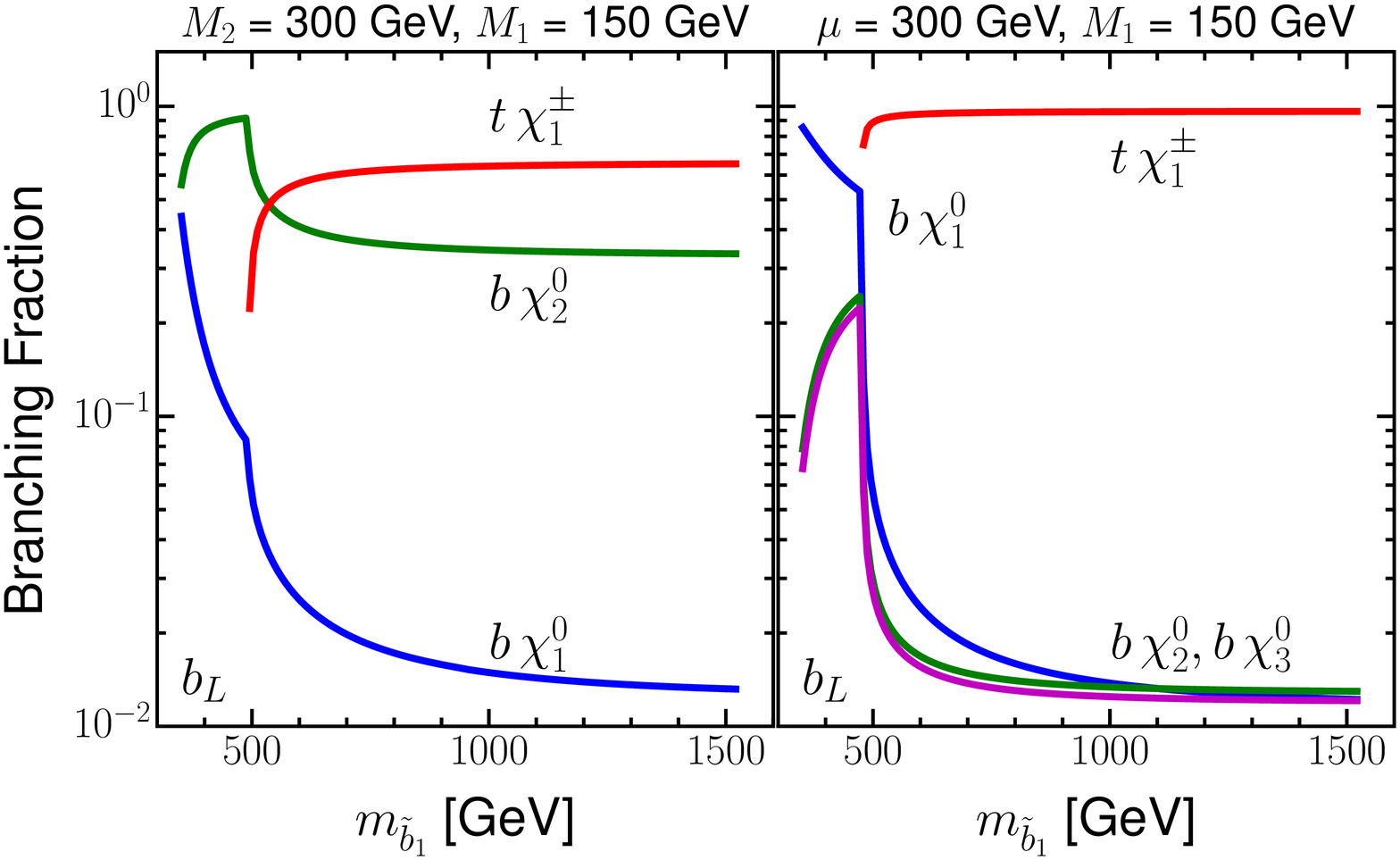}
\minigraph{12.5cm}{-0.3in}{(c)\hspace{2 in}(d)}{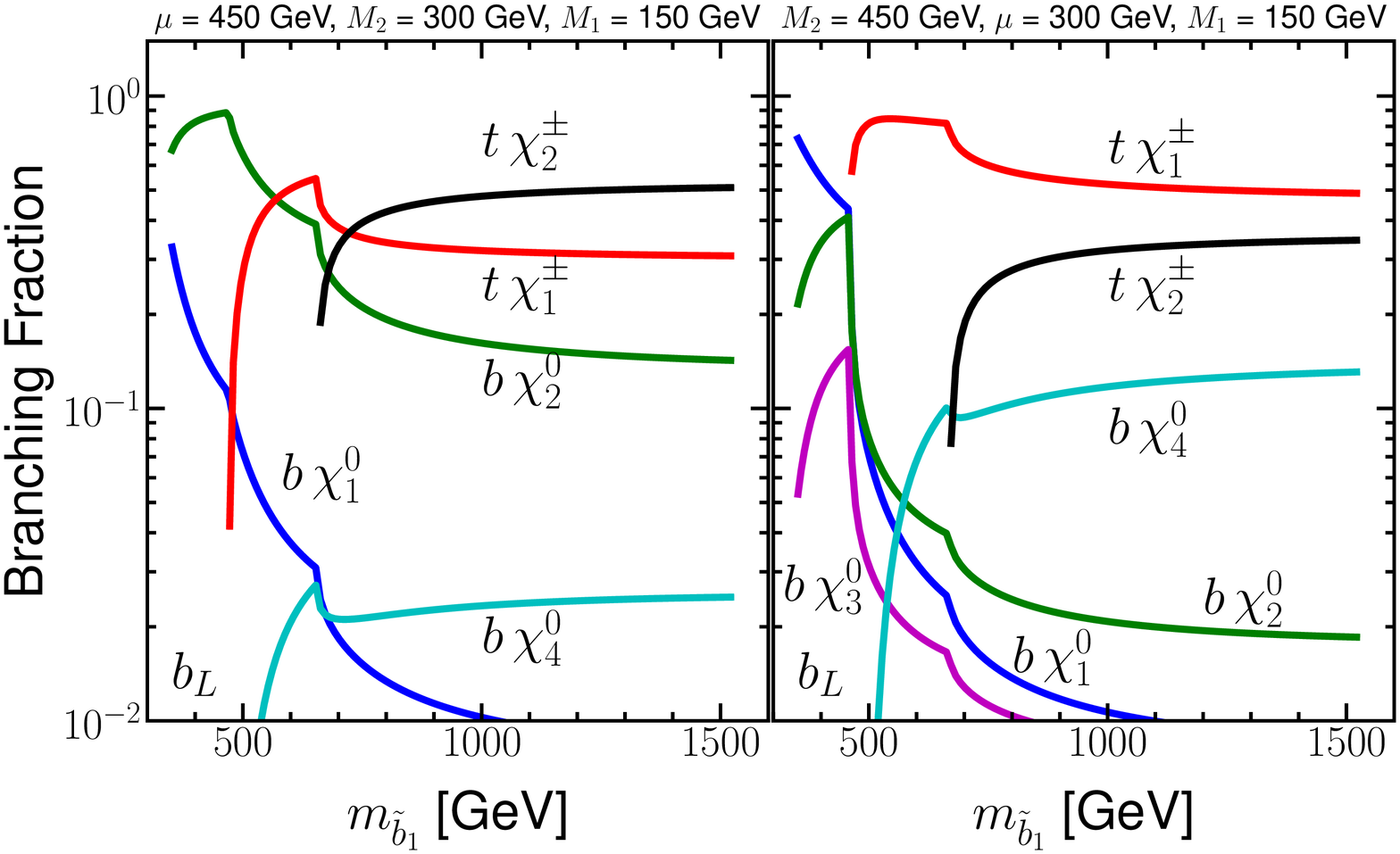}
\caption{ Branch fractions of the left-handed sbottom decay versus its mass in four different cases: (a)   $m_{\tilde{b}_1}>M_2>M_1$, Wino-NLSP; (b)  $m_{\tilde{b}_1}>|\mu|>M_1$, Higgsino-NLSP; (c)  $m_{\tilde{b}_1}>|\mu|>M_2>M_1$, Wino-NLSP/Higgsino-NNLSP, and (d) $m_{\tilde{b}_1}>M_2>|\mu|>M_1$, Higgsino-NLSP/Wino-NNLSP. Here we have adopted $\tan\beta=10$.   } 
\label{fig:BRofsb1}
\end{figure}

We illustrate the sbottom decay in Fig.~\ref{fig:BRofsb1} for these four different situations.
Each corresponds to a different mass spectrum of gaugino and sbottom for a Bino-LSP. 
The usually considered channel $b \chi_1^0$ is suppressed, if other channels are open, since the bino U(1)$_Y$ coupling is smaller than the wino SU(2)$_L$ coupling or top Yukawa coupling. In  Fig.~\ref{fig:BRofsb1}(a), where $\tilde{b}_1\to t \chi_1^\pm$ and $\tilde{b}_1\to b \chi_{2}^0$ are open while the Higgsinos-like neutralinos/charginos are decoupling ($|\mu|>M_{\tilde{b}_1}>M_2>M_1$), sbottom decays dominantly into $ t \chi_1^\pm$ and $b \chi_2^0$. 
Contrarily, in Fig.~\ref{fig:BRofsb1}(b), we decouple Wino-like gaugino while leaving the channel containing Higgsino-like gaugino opening ($M_2>M_{\tilde{b}_1}>|\mu|>M_1$), $\tilde{b}_1\to t \chi_1^\pm $ will soon dominant over other possible channels when the phase space is open due to the large top Yukawa coupling.  $\tilde{b}_1 \rightarrow b \chi_{2,3}^0$ are suppressed due to the relatively small bottom Yukawa coupling. Here we have adopted $\tan\beta=10$. For a larger value of $\tan\beta$, $b \chi_2^0, b \chi_3^0$ channels will be relatively more important.  For more complicated situation, in the lower two panels, we consider the cases of $M_{\tilde{b}_1}>|\mu|>M_2>M_1$ 
(Fig.~\ref{fig:BRofsb1}(c)) and $M_{\tilde{b}_1}>M_2>|\mu|>M_1$ (Fig.~\ref{fig:BRofsb1}(d)). In both cases, sbottom decays dominantly into Higgsino-like chargino, then Wino-like chargino and at last Wino-like neutralino. Other channels are highly suppressed since the U(1)$_Y$ coupling and bottom Yukawa coupling are much smaller. 
% \Shufang{Briefly comment on the case when st1 is lighter than sb1 as well.  This could happen with large At that drag down the lightest stop mass even though sbL and stL share the same M3SQ.}

A special remark is in order. Although $\bL$ and $\sL$ share the same soft mass parameter $M_{3SQ}$, the large mixing between $\sL-\sR$ due to the large trilinear soft SUSY breaking $A_t$ often drags the mass of the (mixed) stop below that of the (mainly left-handed) sbottom. The decay $\tilde{b}_1\rightarrow  W \tilde{t}_1 $ usually dominates once it is kinematically open. However, the about decay patterns still hold as long as $M_{\tilde{b}_1}< M_{\tilde{t}_1} + m_W$. 

%\Tao{The results about are equally applicable for a light stop, as long as it is kinematically forbidden in the decay $\bL \to \tilde{t} +W$.}
%\Shufang{Need to be a bit careful here when applies to stop case, since top and bottom Yukawa coupling relation would be different.  Therefore the relative weights of decays to neutral/charged Higgsino states are different.}
% 
%In Fig.~\ref{fig:BRofsb1FixM3sq}, we also consider two simpler cases where we fix the mass of sbottom, and vary the spectrum of gaugino. In the left panel, the Higgsino-like gaugino are decoupling, as we've discussed above, the dominant channels are $\tilde{b}_1\to\chi_1^-t$ and $\tilde{b}_1\to\chi_2^0b$. In the right panel, we decouple the Wino-like gaugino, $\tilde{b}_1\to\chi_1^-t$ will dominant over all other channel when the phase space is open. The usually mostly considered channel $\tilde{b}_1\to\chi_1^0b$ will dominant only when all other channels have no enough phase space.
%
\begin{figure}[tb]
\minigraph{12.5cm}{-0.3in}{(a)\hspace{2 in}(b)}{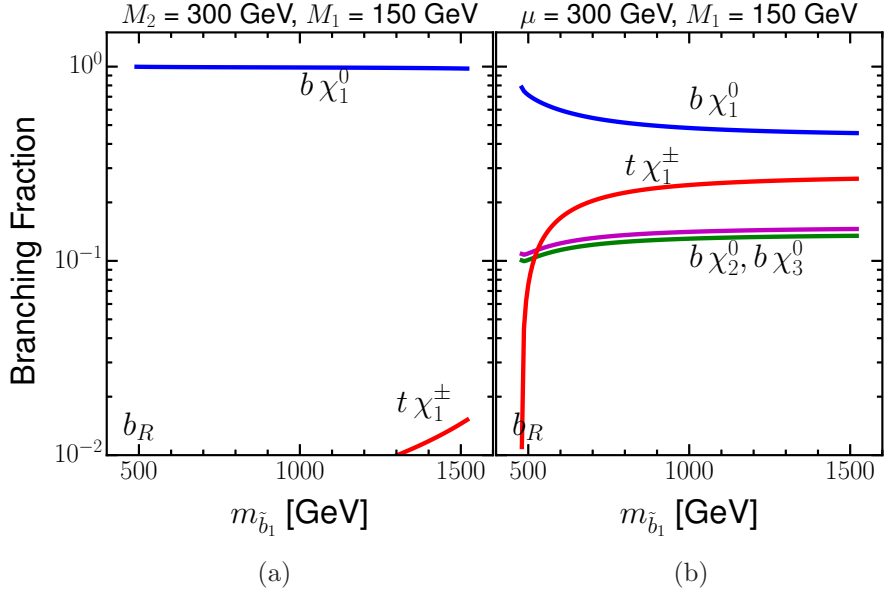}
\caption{ Branch fractions of the right-handed sbottom decay versus its mass for (a)  $m_{\tilde{b}_1}>M_2>M_1$, Wino-NLSP and (b)   $m_{\tilde{b}_1}>|\mu|>M_1$, Higgsino-NLSP.  Here we have adopted $\tan\beta=10$.  }
\label{fig:BRofR}
\end{figure}

\subsubsection{The decay of $\tilde{b}_R$}

For the  $\tilde{b}_R$, the usually considered channel $b \chi_1^0$ is the dominant mode. 
% \Shufang{Again, at large TB, things might change.}  
We present the branching fractions of $\tilde{b}_R$ in Fig.~\ref{fig:BRofR}, for (a) the Wino-NLSP and (b) the Higgsino-NLSP. We see that the channel $\tilde b_1 \to b \chi_1^0$ in the Wino-NLSP scenario is almost 100\%, since the right-handed squark has no SU(2)$_L$ coupling. However, this channel in the Higgsino-NLSP scenario presents a branching fraction about $40\% -60\%$, followed by the channel of $\tilde b_1 \to t \chi_1^\pm$ about $20\% - 30\%$, 
due to the coupling effects of the right-handed squark to Bino or Higgsino is U(1)$_Y$ and bottom Yukawa, respectively.

\subsection{Current bounds from LHC}
\label{sec:limit}

Searches for direct stop and sbottom pair production have been performed at both ATLAS and CMS, with about 5 ${\rm fb}^{-1}$ data at $\sqrt{s}=7$ TeV and 
about 20 ${\rm fb}^{-1}$ data  at $\sqrt{s}=8$ TeV \cite{Aad:2014bva,Aad:2014kra,Aad:2014qaa,CMS:2013nia,CMS-PAS-SUS-14-011,Chatrchyan:2013xna, Aad:2014nra,CMS-PAS-SUS-13-009,Aad:2014mha, Khachatryan:2014doa,CMS-PAS-SUS-13-021,Aad:2013ija,CMS-PAS-SUS-13-018,Aad:2014lra,CMS-PAS-SUS-13-008,Aad:2014pda,Chatrchyan:2013fea,Chatrchyan:2014aea,Chatrchyan:2013mya}. 
The current reach for the stop is slightly worse than that of the sbottom, which has been summarized in Ref.~\cite{Eckel:2014wta}. 
The current searches for the sbottom mainly focus on the decay channel $\tilde{b}_1 \to b \n$ assuming 100\% decay, and the sbottom mass up to 620 (700) GeV is excluded at 95\% C.L. for a massless LSP  with two $b$ plus $\met$ final states based on ATLAS (CMS) analyses \cite{CMS-PAS-SUS-13-018, Aad:2013ija}.  For small mass splitting between sbottom and the LSP: $m_{\tilde{b}}-m_{\chi_1^0}\sim m_b$, monojet plus $\met$ search excludes a sbottom mass up to about 255 GeV~\cite{Aad:2014nra}. 

Sbottom searches for $\tilde{b}\rightarrow b \chi_2^0,\ \chi_2^0\rightarrow \chi_1^0 h$ with 100\% decay branching fraction have been performed at ATLAS~\cite{Aad:2014lra} and the null search results exclude the sbottom masses between 340 and 600 GeV for $m_{\chi_2^0}=300$ GeV and $m_{\chi_1^0}=60$ GeV. For $\tilde{b}\rightarrow b \chi_2^0,\ \chi_2^0\rightarrow \chi_1^0 Z$ with 100\% decay branching fraction, CMS searches exclude sbottom masses up to 450 GeV for LSP masses  between 100 to 125 GeV and $m_{\chi_2^0}-m_{\chi_1^0}=110$ GeV~\cite{CMS-PAS-SUS-13-008}.   Sbottom searches for $\tilde{b}\rightarrow t \chi_1^\pm,\ \chi_1^\pm\rightarrow W \chi_1^0$ with 100\% decay branching fraction have been performed at both ATLAS and CMS~\cite{Aad:2014pda,Chatrchyan:2013fea}.  The sbottom mass limit by ATLAS is about 440 GeV for $m_{\chi_1^\pm} < m_{\tilde{b}}-m_t$~\cite{Aad:2014pda}, and the CMS limits for those channels are about 50 to 100 GeV stronger~\cite{Chatrchyan:2013fea,CMS-PAS-SUS-13-008,Chatrchyan:2014aea}.
We summarize the current search bounds in Table \ref{tab:mbounds}.   

\begin{table}[tb]
\begin{tabular}{|c|c|c|c|}
\toprule[1pt]
Decay channels & Mass bounds $m_{\tilde{b}}$ & BR & Assumptions \\
\midrule[1pt]
%\hline
\multirow{2}{*}{$\tilde{b}_1\to b\n$ (ATLAS \cite{Aad:2013ija})} & 620 GeV & 100\% & $m_{\chi_1^0} <$ 120 GeV \\
& 520 GeV & 60\% & $m_{\chi_1^0} <$ 150 GeV \\ \cline{2-4}
$\tilde{b}_1\to b\n$ (ATLAS \cite{Aad:2014nra}) & 255 GeV & 100\% & $m_{\tilde{b}}-m_{\chi_1^0}\sim m_b$ \\
$\tilde{b}_1\to b\n$ (CMS \cite{CMS-PAS-SUS-13-018}) & 700 GeV & 100\% & Small $m_{\chi_1^0}$ \\
\hline
\multirow{2}{*}{$\tilde{b}_1\to b\nn\to b h\n$ (ATLAS \cite{Aad:2014lra})} & \multirow{2}{*}{340 - 600 GeV} & \multirow{2}{*}{100\%} & $m_{\chi_2^0}$ = 300 GeV \\
& & & $m_{\chi_1^0}$ = 60 GeV \\ \cline{2-4}
%$\tilde{b}_1\to b\nn\to b Z\n$ (ATLAS \cite{})& 440 GeV, 50 GeV & 100\% & $m_{\chi_1^\pm} < m_{\tilde{b}}-m_t$ \\
%$\tilde{b}_1\to b\nn\to b h\n$ (CMS) & 440 GeV, 50 GeV& 100\% & $m_{\chi_1^\pm} < m_{\tilde{b}}-m_t$ \\
\multirow{2}{*}{$\tilde{b}_1\to b\nn\to b Z\n$ (CMS \cite{CMS-PAS-SUS-13-008})} & \multirow{2}{*}{450 GeV} & \multirow{2}{*}{100\%} & 100 GeV $< m_{\chi_1^0} <$ 125 GeV \\
& & & $m_{\chi_2^0} - m_{\chi_1^0}$ = 110 GeV \\
\hline
$\tilde{b}_1\to t\chi_1^-$ (ATLAS \cite{Aad:2014pda}) & 440 GeV & 100\% & $m_{\chi_1^\pm} < m_{\tilde{b}}-m_t$ \\ \cline{2-4}
\multirow{6}{*}{$\tilde{b}_1\to t\chi_1^-$ (CMS \cite{CMS-PAS-SUS-13-008})} & \multirow{2}{*}{575 GeV}  & \multirow{2}{*}{100\%} & 150 GeV $< m_{\chi_1^\pm} <$ 375 GeV \\
& & & $m_{\chi_1^0}$ = 50 GeV \\ \cline{2-4}
& \multirow{2}{*}{575 GeV} & \multirow{2}{*}{100\%} & 25 GeV $< m_{\chi_1^0} <$ 150 GeV \\
& & & $\frac{m_{\chi_1^0}}{m_{\chi_1^\pm}}$ = 0.5 \\ \cline{2-4}
& \multirow{2}{*}{525 GeV} & \multirow{2}{*}{100\%} & 25 GeV $< m_{\chi_1^0} <$ 200 GeV \\
& & & $\frac{m_{\chi_1^0}}{m_{\chi_1^\pm}}$ = 0.8 \\ \cline{2-4}
$\tilde{b}_1\to t\chi_1^-$ (CMS \cite{Chatrchyan:2013fea}) & 500 GeV & 100\% & $\frac{m_{\chi_1^0}}{m_{\chi_1^\pm}}$ = 0.5 (0.8) \\
$\tilde{b}_1\to t\chi_1^-$ (CMS \cite{Chatrchyan:2014aea}) & 550 GeV & 100\% & $m_{\chi_1^0}$ = 50 GeV \\
\bottomrule[1pt]
\end{tabular}
\caption{Current mass bounds on the sbottom from the direct searches at the LHC.  } 
\label{tab:mbounds}
\end{table}

%%%%%%%%%%%%%%%%%%%%%%%%%%%%%%%%%%%%%%%%%%%

\section{LHC analysis}
\label{sec:analyses}
In this section, we study the collider phenomenology of the light sbottom at the 14 TeV LHC. The key point in this paper is to explore the mixed decay channels according to the mass hierarchies
beyond the common assumption of $100\%$ branching fraction of a given channel.
%
% \Yongcheng{Further, we do not consider the non-mixed channel, e.g. $\tilde{b}_1\tilde{b}_1^*\to b\chi_2^0b\chi_2^0$ or $\tilde{b}_1\tilde{b}_1^*\to t\chi_1^\pm t\chi_1^\mp$, for these channels have already been investigated by Experiments, as listed in Tab.\ref{tab:mbounds}. Our cases can offer no more interesting phenomenon besides the different branch faction compared with the result from Experiments. We will just give the relaxed bounds(if exist) for these channels using more realistic branch fraction. } We first study the left-handed sbottom since the decay pattern is richer. We then show the results for the right-handed sbottom. 
%
Including those channels listed in Table I with realistic branching fractions would help increase the overall sensitivity, but we did not repeat the analyses.   We note that Ref.~\cite{Graesser:2012qy} also exploited the mixed decays to search for stop. They introduced a new variable ``topness" for the top-rich signal events to help efficiently reduce the top pair backgrounds.

\subsection{Signature of $\tilde{b}_1 \sim \bL$}
\label{sec:LHsbottom}
 We consider the scenario with the low energy mass spectrum containing a light sbottom (mostly left-handed), a Bino-like LSP and Wino-like NLSPs, as shown in Fig.~\ref{fig:BRofsb1}(a). Two typical benchmark points  for both signs of $\mu$ are listed in Table.~\ref{tab:massParameters}. Other soft SUSY breaking parameters are decoupled to be 2 TeV, and $\tilde A_t$ is set to be large such that the SM-like Higgs is around 125 \gev.   The value of $\mu$ is chosen such that $\chi_2^0$ dominantly decays to $h \chi_1^0$ for $\mu>0$ and to $Z\chi_1^0$ for $\mu<0$\footnote{Note that  $\chi_2^0\rightarrow Z\chi_1^0$ is not always dominated for $\mu<0$, as pointed out in Refs.~\cite{gunion1988two, Jung:2014bda}.  We have chosen the value of $\mu$ in the $\mu<0$ case  to guarantee the $Z$ channel dominance.}.
The decay channels and the corresponding decay branching fractions for $\tilde{b}_1$, $\tilde{t}_1$, as well as $\chi_2^0$ and $\chi_1^\pm$ are listed in Table.~\ref{tab:DecayBR}.   The  conventional channel $\tilde{b}_1 \to b \n$ is highly suppressed,  with only about 2\% branching fraction, which dramatically weakens the current experimental search limit.   The decay channels of $\tilde{b}_1 \to b \nn$ and $\tilde{b}_1 \to t \chi^-_1$ are comparable and  dominant instead.     In particular, with one sbottom decaying to $\chi_2^0$ and one sbottom decaying to $\chi_1^\pm$, $\tilde{b}_1\tilde{b}_1^*$ pair production leads to interesting final states of $bbWW+h/Z+\met$.   Note that unmixed decays of $\tilde{b}_1\tilde{b}_1^* \rightarrow bbhh+\met$, $bbZZ+\met$, $ttWW+\met$ have been studied at the LHC~\cite{Aad:2014lra,CMS-PAS-SUS-13-008,Aad:2014pda,Chatrchyan:2013fea,Chatrchyan:2014aea}, assuming 100\% decay branching fractions.  Given the more realistic branching fractions  of about 40\% for $\tilde{b}_1\rightarrow b\chi_2^0$ and about 60\% for  $\tilde{b}_1\rightarrow t\chi_1^-$, the collider limits for those channels will be relaxed.  Including all the mixed and unmixed channels can further increase the collider reach for the sbottom.

\begin{table}[tb]
\begin{tabular}{|c|c|c|c|c|c|c||c|c|c|c|c|c|}
\hline
&$M_1$ & $M_2$ & $M_{3SQ}$& $A_t$ & $\mu$ & $\tan\beta$ & $m_{\chi_1^0}$ & $m_{\chi_2^0}$ & $m_{\chi_1^\pm}$ &$m_{\tilde{b}_1}$ & $m_{\tilde{t}_1}$ & $m_h$ \\
\hline
%150 & 300 & 550 & 2950 & +2000 & 10 &151 & 319 & 319 & 526 & 538 & 125 \\
BP1&150 & 300 & 650 & 2950 & +2000 & 10 & 152 & 320 & 320 & 640 & 650 & 125 \\
\hline
%150 & 300 & 550 & 2950 & $-$1300 & 10 & 152 & 322 & 321 & 526 & 525 & 124 \\
BP2&150 & 300 & 650 & 2950 & $-$1300 & 10 & 150 & 320 & 320 & 640 & 630 & 125 \\
\hline
\end{tabular}
\caption{MSSM parameters and mass spectrum of SUSY particles for the two benchmark points.   All masses are in units of GeV.}
%\Shufang{Can we use a benchmark point with stop/sbottom mass of 650 GeV?  While I think the current value is still safe, it is on the border line given the current LHC stop/sbottom search limit.  Let's use 650 GeV to be safe.   Table I-IV and Fig 1,3 need to be changed.  Thanks! }\Yongcheng{This table has been changed to 650GeV-case!} \Yongcheng{Tables and Figures have all been changed to 650GeV case.}}
\label{tab:massParameters}
\end{table}

\begin{table}[tb]
\begin{tabular}{|c|c|c||c|c||c|c|}
\toprule[1pt]
&Decay Channel & BR & Decay Channel & BR & Decay Channel & BR \\
\midrule[1pt]
\multirow{3}{*}{BP1 ($\mu>0$)}&$\tilde{b}_1\to b\n$ & 2\% & $\tilde{t}_1 \to t \n$ & 2\% & $\nn\to h\n$ & 97\% \\
\cline{2-7}
&$\tilde{b}_1\to b\nn$ & 39\% & $\tilde{t}_1 \to t \nn$ & 27\% & $\nn\to Z\n$ & 3\% \\
\cline{2-7}
&$\tilde{b}_1\to t\chi_1^-$ & 59\% & $\tilde{t}_1 \to b \cha$ & 71\% & $\chi_1^\pm\to W^\pm\n$ & 100\% \\
\midrule[1pt]
\multirow{3}{*}{BP2 ($\mu<0$)}&$\tilde{b}_1\to b\n$ & 2\% & $\tilde{t}_1 \to t \n$ & 2\% & $\nn\to h\n$ & 6\% \\
\cline{2-7}
&$\tilde{b}_1\to b\nn$ & 39\% & $\tilde{t}_1 \to t \nn$ & 27 \% & $\nn\to Z\n$ & 94\% \\
\cline{2-7}
&$\tilde{b}_1\to t\chi_1^-$ & 59\% & $\tilde{t}_1 \to b \cha$ & 71\% & $\chi_1^\pm\to W^\pm\n$ & 100\% \\
\bottomrule[1pt]
\end{tabular}
\caption{Decay channels and the corresponding branching fractions of $\tilde{b}_1$, $\tilde{t}_1$, $\nn$ and $\cha$ for the   two benchmark points, which correspond to the cases of $\mu>0$ and $\mu<0$. } % \Shufang{ Do we need the table? $\mu>0$ and $\mu<0$ are very similar for stop and sbottom decay, with the only difference being chi20 decay.}}
\label{tab:DecayBR}
\end{table}

The stop decay  has been   studied in detail in Ref.~\cite{Eckel:2014wta}.  For the two benchmark points listed in Table.~\ref{tab:massParameters},  the conventional decay channel $\tilde{t}_1 \to t \n$ is highly suppressed.  $\tilde{t}_1 \to b \chi^-_1$ is dominant with branching fraction of about 70\%.  $\tilde{t}_1 \to t \nn$ is subdominant with a branching fraction of about 27\%.   With one stop decaying to $\chi_2^0$ and one stop decaying to $\chi_1^0$, $\tilde{t}_1\tilde{t}_1^*$ pair production provides the same final states as the sbottom case.  

%, due to the low $U(1)_Y$ coupling compared to the relative high top Yukawa coupling and $U(2)_L$ coupling. And the decay channel of $\tilde{t}_1 \to b \chi^-_1$ is dominant, about 78\% branching ratio, and the subdominant decay channel is $\tilde{t}_1 \to t \n$, with branching ratio of 20\%, providing the same final states as the light sbottom case where $\nn$ decays dominantly to $\n h$ with 97\% branching ratio. 

The two benchmark points listed in Table.~\ref{tab:massParameters} are only for illustration whenever instructive. In our following analyses, we perform a broad scan over the mass parameter space.
 
$\bullet$ 
$M_{3SQ}$ from 400 to 1075 GeV with a step size of 25 GeV, corresponding to $m_{\tilde{b}_1}$ from about 350 GeV to about 1085 GeV and $m_{\tilde{t}_1}$ from about 367 GeV to about 1090 GeV.  

$\bullet$ 
$M_1$ is scanned from 3 GeV to 700 GeV,  in the step of 25 GeV. 

$\bullet$ 
$M_2$ is fixed to be $M_2=M_1+150~\gev$. 

$\bullet$ 
We further require $m_{\tilde{b}_1}>m_{\chi_1^\pm}+m_t$ such that $\tilde{b}_1\to t\chi_1^\pm$  can be open. \\
 
%Further, we've also required $m_{\tilde{b}_1}>m_{\chi_1^\pm}+m_t+10\approx m_{\chi_1^\pm}+185$ such that the channel $\tilde{b}_1\to t\chi_1^\pm$ can open. Hence, we get the limitation for our parameter space of $m_{\tilde{b}_1}-m_{\chi_1^0}$ plane (in $\gev$):
%\begin{equation}
%\label{equ:LimitationI}
%m_{\tilde{b}_1}>m_{\n}+335.
%\end{equation}
%
%Under these limits, we've scanned 425 points in the parameter space and generated events for each points at $\sqrt{s}=14\text{ TeV}$. 

%The decay channels of related particles and corresponding branching ratio are also shown in Table.~\ref{tab:DecayBR}. The decay patterns of the light stop and the light sbottom are barely affected by the choice sign of $\mu$, the significant difference lies in the $\nn$ decay, here $\nn$ dominantly decays to $\n Z$, compared to the $h$ channel for $\mu >$ 0 case.

%\begin{table}[!htb]
%\begin{tabular}{|c|c||c|c|}
%\hline
%Decay Channel & Branching Ratio & Decay Channel & Branching Ratio \\
%\hline
%$\tilde{b}_1\to b\n$ & 3.84\% & $\nn\to h\n$ & 0.56\% \\
%\hline
%$\tilde{b}_1\to b\nn$ & 51.20\% & $\nn\to Z\n$ & 99.44\% \\
%\hline
%$\tilde{b}_1\to t\chi_1^-$ & 44.96\% & $\chi_1^\pm\to W^\pm\n$ & 100\% \\
%\hline
%\end{tabular}
%\caption{Decay Channels and Corresponding Branching Ratio of Relevant Particles for the Case of $\mu<0$. }
%\label{tab:minusmuBR}
%\end{table}

In our phenomenological studies, we define the basic observable objects as  
\begin{itemize}
\item Jet:
\begin{equation}
\label{equ:BasicCutsJets}
|\eta_{j}|<2.5 ,\quad p_{T}^{j} > 25 \text{ GeV} ,\quad \Delta\phi_{j,\met} > 0.8.
\end{equation}
where $\Delta\phi_{j,\slashed{E}_T}$ is azimuthal angle between the jet and missing transverse energy.
\item  Lepton:
\begin{equation}
\label{equ:BasicCutsLep}
|\eta_{\ell}|<2.5 ,\quad p_{T}^{\ell} > 20 \text{ GeV} ,\quad \Delta R_{\ell j} > 0.4.
\end{equation}
Where the $\Delta R_{\ell j}$ is the distance in the $\phi$-$\eta$ plane: $\Delta R = \sqrt{\Delta\phi^2+\Delta\eta^2}$,  between the lepton and the jet satisfying Eq.~(\ref{equ:BasicCutsJets}).
\end{itemize}
To be as realistic as possible, both the signal and the background samples are generated by MadGraph~5 \cite{MADGRAPH5V2}, passed through Pythia~6~\cite{PYTHIA6} for the fragmentation and hadronization. We further perform the detector simulation through Delphes~3~\cite{DELPHES3} with Snowmass Delphes No-Pile-up detector cards~\cite{Anderson:2013kxz}.

\subsubsection{The Case of $\mu>0$: final states with a Higgs}
\label{subsec:plusmu}

%Combining all these decay channels, we list the SUSY process we are interested in and corresponding cross section in the Table.~\ref{tab:plusmuProcess}. For the SUSY signal process, we considered sbottom production under 14 TeV at LHC, with one sbottom decaying via $\tilde{b}_1\to b\nn\to b h\n$, the other one decaying via $\tilde{b}_1\to t\chi^\pm_1\to bW^\pm W^\mp\n$. Among these two W-boson, we have one decaying leptonically and the other decaying hadronically. And the higgs decays into two bottom quark. Hence in the signal process, we have 4 b-jets, 2 light jets, 1 lepton and missing energy. The corresponding SM background processes are listed in the Table.~\ref{tab:plusmuProcess} as well.  Among these background processes, the $t\bar{t}$ and $t\bar{t}b\bar{b}$ are the dominant processes. We also considered the production of top-quark pair associated with gauge boson or higgs $t\bar{t}Z$, $t\bar{t}W^\pm$ and $t\bar{t}h$. The cross section in the Table.~\ref{tab:plusmuProcess} for each process is normalized to NLO value according to the calculation in several papers~\cite{stopkfactor,3rdsquarkkfactor,NLOtt,NLOttbb,NLOttz,NLOtth,NLOttw}.
 
In the case of $\mu>0$, the leading signal under consideration for the pair production of sbottom, with $\tilde{b}_1 \to b \nn \to b h \n$ and $\tilde{b}^*_1 \to t\chi_1^-\to bW^+\ W^-\n$, is 
$$\tilde{b}_1\tilde{b}^*_1 \rightarrow bb\ WW\ h\ \met \rightarrow \ell\ bbbb\ jj \met.$$  
The signal contains four $b$-jets, two light flavor jets, one isolated lepton ($e$ or $\mu$), and large missing energy. 
The study of the same final state from stop decay can be found in Ref.~\cite{Eckel:2014wta}. 
%
%The existence of the lepton will significantly reduce the QCD backgrounds without much suppresion of the branching ratio. 
The dominant backgrounds will be from $t \bar t+$jets and $t \bar t b \bar b$ with large cross sections and similar final states. While $t \bar t h$ is an irreducible background, the production cross section is relatively small. Other SM backgrounds include $t \bar t W$, $t \bar t Z$ and $b \bar b W W$, with typically smaller cross sections.  

\begin{comment}
 \begin{table}[tb]
\centering
\begin{tabular}{|l|l|l|}
\hline
& Processes & $\sigma$ \\
\hline
Signal&$p\ p\ \to\ \tilde{b}_1\ \bar{\tilde{b}}_1 \to\ b\ h\ t\ W\ \slashed{E}_T\to\ 4b+2jets+1lep+\slashed{E}_T$ & 43.13fb\\
\hline
\multirow{5}{*}{Background} &$p\ p\ \to\ t\ \bar{t}\to\ 2b+2jet+1lep+\slashed{E}_T$ & 261.23pb\\

&$p\ p\ \to\ t\ \bar{t}\ b\ \bar{b}\to\ 4b+2jet+1lep+\slashed{E}_T$ & 2.330pb \\

&$p\ p\ \to\ t\ \bar{t}\ Z\to\ 4b+2jet+1lep+\slashed{E}_T$ & 228.177fb\\

&$p\ p\ \to\ t\ \bar{t}\ h\to\ 4b+2jet+1lep+\slashed{E}_T$ & 102.08fb\\

&$p\ p\ \to\ t\ \bar{t}\ W^{\pm}\to\ 2b+4jet+1lep+\slashed{E}_T$ & 223.84fb \\

\hline
\end{tabular}
\caption{Signal and Background Processes for the Case of $\mu>0$, the Cross Sections are calculated under 14 TeV.}
\label{tab:plusmuProcess}
\end{table}
\end{comment}

To select the signal of $\tilde{b}_1\tilde{b}_1^*,\ \tilde{t}_1\tilde{t}_1^* \rightarrow bb\ WW\ h\ \met \rightarrow \ell\ bbbb\ jj\ \met$, we adopt the basic event selection 
\begin{itemize}
\item   $N_j \geq 4, \ p_{T}^{j1,j2,j3} > 40 \text{ GeV}, \  N_\ell = 1$.    
 \end{itemize}
Beside these basic cuts, we further optimize the cuts and divide the events into signal regions on the following variables: 
\begin{itemize}
\item Missing energy $\met$,  which is the magnitude of the the missing transverse momentum, to be above
100, 120, 140, 160 180, and 200 GeV.
 \item $H_T$, the scalar sum of the jet transverse momentum of all surviving isolated jets:
$H_T = \sum_{\text{jets}} |p_T^j|$, 
to be above 400, 450, 500, 550, 600 GeV.
\begin{comment}
\item $M_{\text{eff}}$, the effective mass of event, defined as scalar sum of transverse momentum $p_T$ of all visible final states in addition with Missing Energy:
\begin{equation}
M_{\text{eff}}=\sum_{\text{all visible particles}}|p_T| + \slashed{E}_T
\end{equation}
\end{comment}
 \item $M_T$, the transverse mass, defined as the invariant mass of the lepton and missing energy:
\begin{equation}
\label{equ:MT}
M_T({\bf p}_T^\ell,{\bf p}_T^{\rm miss})=\sqrt{2p_T^\ell p_T^{\rm miss}(1-\cos\phi_{\ell,\met})},
\end{equation} 
to be above 100, 120, 140, 160, 180, 200 GeV.   
\item $N_j$, the multiplicity of all surviving isolated jets, being at least 4, 5 and 6. 
\item $N_b$, the multiplicity of tagged $b$-jets,  being at least 2, 3 and 4.  
\end{itemize} 

\begin{figure}[tb]
%
%\subfloat[$M_{\text{eff}}$]{
%\begin{minipage}[t]{0.5\textwidth}
%\includegraphics[width = 0.45 \textwidth]{./Figures/CasePlusMu/Meff_all.eps}
%\end{minipage}
%}
%\subfloat[$H_T$]{
%\begin{minipage}[t]{0.5\textwidth}
%\includegraphics[width = 0.49 \textwidth]{./Figures/CasePlusMu/HT_all.eps}
%\end{minipage}
%}
%
%\subfloat[$\slashed{E}_T$]{
%\begin{minipage}[t]{0.5\textwidth}
\includegraphics[width = 0.49 \textwidth]{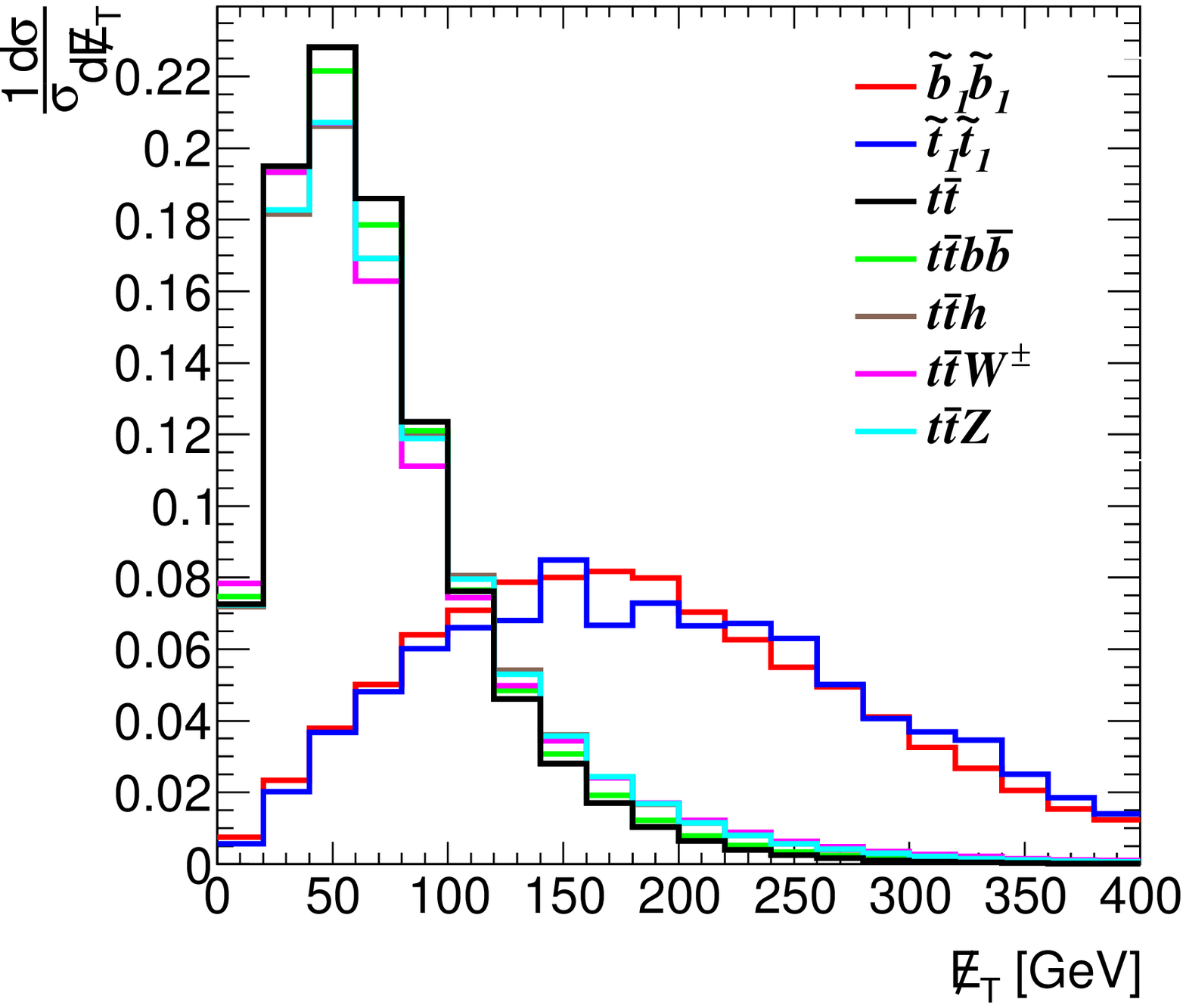}
%\end{minipage}
%}
%\subfloat[$M_T$]{
%\begin{minipage}[t]{0.5\textwidth}
%\includegraphics[width = 0.49 \textwidth]{./Figures/CasePlusMu/MT_all_withstop_650.eps}
\includegraphics[width = 0.49 \textwidth]{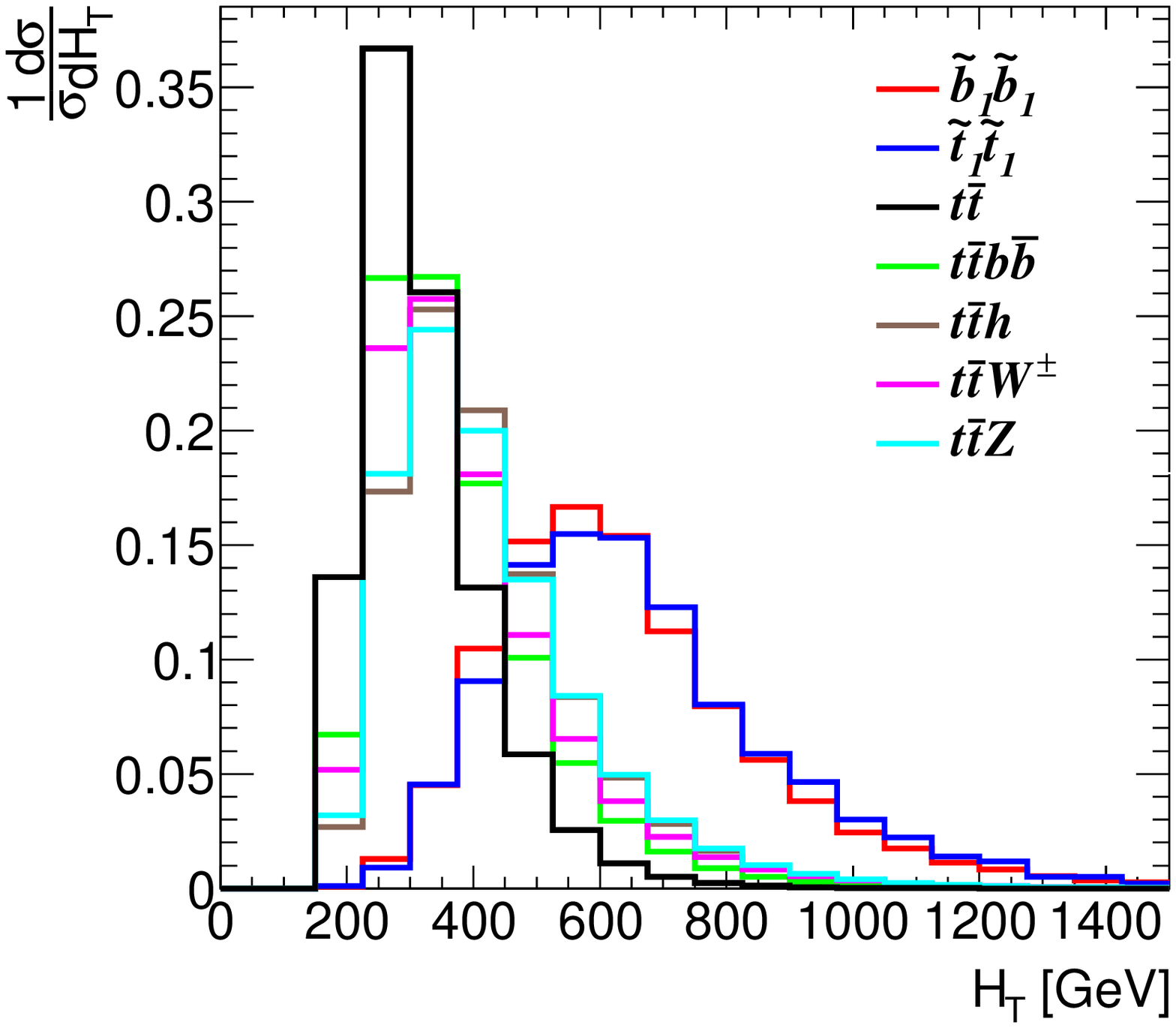}
%\end{minipage}
%}
\caption{Normalized distributions of $\met$ (left panel) and 
%$M_T$ (left panel) and 
$H_T$ (right panel) for the signal $\tilde{b}_1\tilde{b}_1^*$ (red curves), $\tilde{t}_1\tilde{t}_1^*$ (blue curves)  $\rightarrow bbWWh\met\rightarrow  \ell\ bbbb\ jj\  \met$ after basic cuts with $m_{\tilde{b}_1}=637$ GeV, $m_{\tilde{t}_1}=646$ GeV, as well as SM backgrounds at the 14 TeV LHC.}
\label{fig:VariablesI}
\end{figure}

The normalized distributions of  $\slashed{E}_T$ and $H_T$ are shown in Fig.~\ref{fig:VariablesI}. As expected,  the signal process has larger $\slashed{E}_T$ from the missing neutralino-LSP than the background processes, which is typically bounded by 
%The $M_T$ distributions for the SM backgrounds have a sharp edge around 
$m_W/2$ due to the primary contribution $W \rightarrow \ell \nu$.
%while the $M_T$ distribution for the signal extends to large value.    
Given the relatively large sbottom mass,  the signal process typically has larger $H_T$ than the SM backgrounds as well. 

In Table.~\ref{tab:CUTSResultI}, we list the cumulative cut efficiencies after different levels of cuts, as well as cross sections before and after cuts for both the  sbottom and stop signals as well as the SM backgrounds for the benchmark point listed in Table \ref{tab:massParameters} for $\mu > 0$.   The cross section for each process is normalized to their theoretical values including NLO QCD corrections \cite{stopkfactor,3rdsquarkkfactor,NLOtt,NLOttbb,NLOttz,NLOtth,NLOttw}. The background processes are significantly suppressed after strong $\met$, $H_T$, $M_T$ cuts. The leading background left is $t\bar{t}$, followed by $t\bar{t}b\bar{b}$. We scan over the combinations of the signal regions, to select the optimal combination which gives the best significance for each mass grid point, including 10\% systematic uncertainty.   At $\sqrt{s}=14$ TeV with $300 \ {\rm fb}^{-1}$ integrated luminosity, the significance could reach about  $17\sigma$ ($14\sigma$) for $\tilde b_1$ \ ($\tilde t_1$) of about 640 GeV.  

\begin{table}[tb]
\begin{tabular}{|c|c|c|c|c|c|c|c|c|}%{!{\vrule width 1pt}r!{\vrule width 1pt}r|r|r|r|r|r|r|r!{\vrule width 1pt}}%%
\toprule[1pt]
%\hline
Process & $\sigma$ (fb)& Basic &$\slashed{E}_T>$ & $H_T>$ & $M_T>$ & $N_j\geq$ & $N_b\geq$ & $\sigma$ (fb)\\
        &  & cuts & $200\text{ GeV}$ & $500\text{ GeV}$ & $160\text{ GeV}$ & $5$ & $2$ & after cuts \\ 
\toprule[1pt]
%\hline
%$\tilde{b}_1 \tilde{b}_1$ & 43.13 & 35\% & 8.7\% & 5.3\% & 1.9\% & 1.5\% & 0.80\% & 0.34 \\
$\tilde{b}_1 \tilde{b}_1$ & 13 & 39\% & 17\% & 14\% & 5.8\% & 4.3\% & 2.7\% & $3.4 \times 10^{-1}$ \\
\hline
%$\tilde{t}_1 \tilde{t}_1$ & 23 & 38\% & 12\% & 7.6\% & 2.4\% & 2.0\% & 1.2\% & 0.28 \\ 
$\tilde{t}_1 \tilde{t}_1$ & 10 & 39\% & 18\% & 16\% & 5.9\% & 4.4\% & 2.9\% & $2.9 \times 10^{-1}$ \\ 
\hline
$t\bar{t}$ & 260,000 &  14\% &  0.24\% &  $7.4\times10^{-4}$ & $1.7\times10^{-6}$ & $9.3\times10^{-7}$ & $2.4\times10^{-7}$ & $6.3\times 10^{-2}$ \\
\hline
$t\bar{t}b\bar{b}$  & 2,300& 24\% & 0.6\%  & 0.3\% & $3.5\times10^{-5}$ & $2.3\times10^{-5}$ & $1.2\times10^{-5}$ &$ 2.8 \times 10^{-2}$\\
\hline
$t\bar{t}h$ & 100 & 31\% & 1.2\% & 0.8\% & $5.8\times10^{-5}$ & $3.4\times10^{-5}$ & $1.9\times10^{-5}$ & $2.0 \times 10^{-3}$\\
\hline
$t\bar{t}Z$ & 230 & 30\% & 1.2\% & 0.8\% & $6.6\times10^{-5}$ & $3.9\times10^{-5}$ & $9.8\times10^{-6}$ &$2.2 \times 10^{-3}$\\
\hline
$t\bar{t}W^{\pm}$ &224 & 25\% & 1.2\% & 0.7\% & $4.8\times10^{-5}$ & $2.3\times10^{-5} $ & $6.3\times10^{-6}$ & $1.4 \times 10^{-3}$\\
\hline
 & &\multicolumn{5}{r}{$\sqrt{s}=14\ {\rm TeV} \qquad \int L \ dt = 300 \ {\rm fb}^{-1} \qquad$} &  &  \\
 & & \multicolumn{5}{c}{${S\over\sqrt{B+(10\%B)^2}}$ = 17 \ (14) \quad for \quad $\tilde b_1$ \ ($\tilde t_1$)} &  &  \\
\bottomrule[1pt]
\hline
\end{tabular}
\caption{Cut efficiencies and cross sections before and after cuts for the signal $\tilde{b}_1\tilde{b}^*_1, \tilde{t}_1\tilde{t}_1^* \rightarrow bbWWh\met \rightarrow \ell\  bbbb\ jj\ \met$ for BP1 listed in Table \ref{tab:massParameters} for $\mu > 0$, as well as SM backgrounds at the 14 TeV LHC.  The significance is  obtained  for  $\int L dt$ = 300 ${\rm fb}^{-1}$ with 10\% systematic error combining both sbottom and stop signals.   }
 \label{tab:CUTSResultI}
\end{table}

\begin{figure}[tb]
\minigraph{7cm}{-0.3in}{(a)}{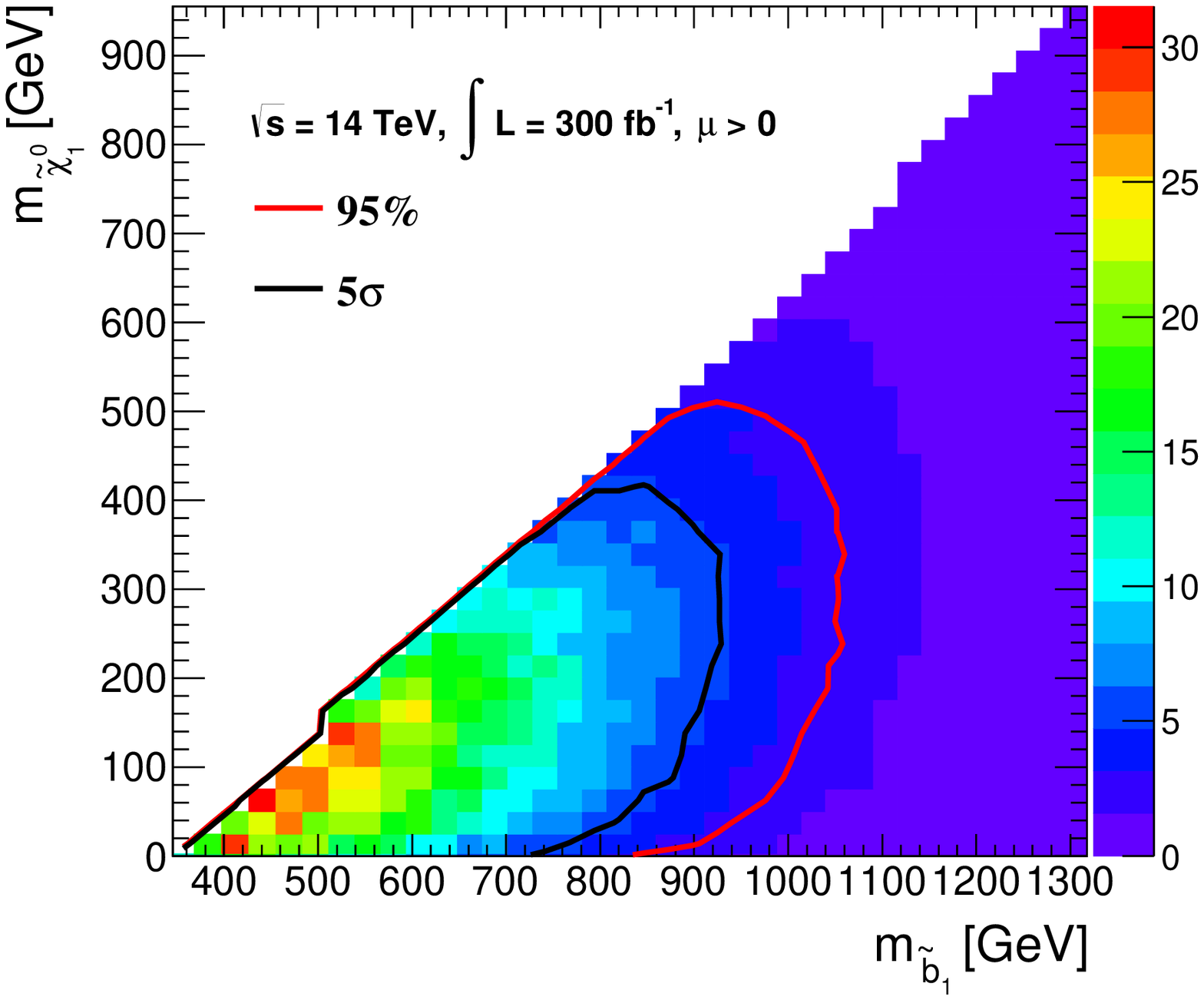}
\minigraph{7cm}{-0.3in}{(b)}{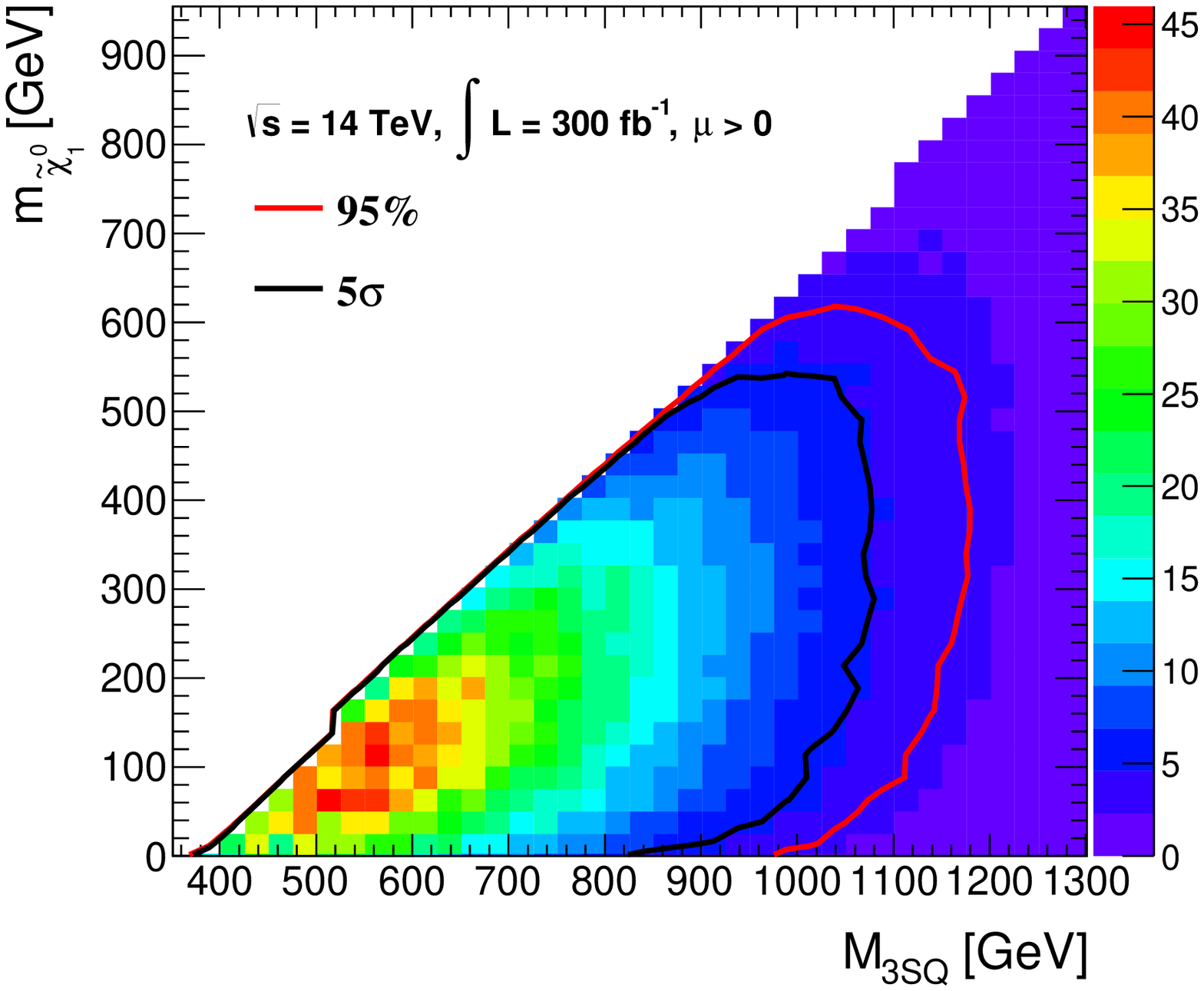}
%\includegraphics[scale=0.40]{./Figures/CasePlusMu/PlusMu_SGCL_Contour}
%\ includegraphics[scale=0.40]{./Figures/CasePlusMu/combine_M3SQ}
\caption{
Signal significance contours for $\tilde{b}_1\tilde{b}^*, \tilde{t}_1\tilde{t}_1^* \rightarrow bbWWh\met \rightarrow \ell\ bbbb\ jj\ \met$ final states for 14 TeV LHC with ${\int L \ dt}\ = \ 300\text{ fb}^{-1}$ luminosity. The 5$\sigma$ discovery reach (black curves) and 95\% C.L exclusion limit (red curves) for the sbottom only are shown in the (a) $m_{\tilde{b}_1}-m_{\chi_1^0}$ plane, and in the (b) $M_{3SQ} - m_{\chi_1^0}$ plane  for 
% the reach for stop only is shown in Figure 13 in Re.~\cite{Eckel:2014wta} and 
the combined reach for sbottom and stop. }
\label{fig:SGCLI}
\end{figure}

Signal significance contours are shown in Fig.~\ref{fig:SGCLI} with the 5$\sigma$ discovery reach (black curve) and 95\% C.L. exclusion limit (red curve) for 14 TeV LHC with 300 ${\rm fb}^{-1}$ integrated luminosity. Fig.~\ref{fig:SGCLI} (a)  shows the $m_{\tilde{b}_1}-m_{\chi_1^0}$ plane.
We find that 5$\sigma$ discovery can reach about 750~$\gev$ for $\tilde{b}_1$ when $\n$ is almost massless and reach about 920~$\gev$ when $\n$ is about 200~$\gev$ to 300~$\gev$.  The 95\% C.L.  exclusion reach is about 100 GeV better.   The reach for the stop with the same final states can be found in Ref.~\cite{Eckel:2014wta}, with results being very similar.  

Since the (mostly left-handed) sbottom and stop have  the same undistinguishable final states with their masses controlled by the same parameter $M_{3SQ}$,  we present the combined reach of stop and sbottom in  Fig.~\ref{fig:SGCLI} (b) in $M_{3SQ}-m_{\chi_1^0}$ plane\footnote{The mass difference between the stop and sbottom  does not affect the combination of the stop and sbottom signals, since the same cuts are used for both the stop and sbottom events.}.  The 5$\sigma$ discovery reach in $M_{3SQ}$ increases to be 820 GeV for a massless LSP, and 1080 GeV for $m_{\n} \sim$ 300 GeV. The masses up to 980 GeV can be excluded for a massless LSP, and the masses up to 1180 GeV can be excluded for $m_{\n} \sim$ 300 GeV at 95\% C.L.   

We would like to reiterate that the mixing in sbottom and stop sectors governs the mass spectrum of the sbottom and stop.
Small mixing in the sbottom sector is always a good approximation given the small bottom Yukawa coupling, while the mixing in the stop sector may be large enough to suppress the mass of the lighter stop further. In our cases (including the $\mu<0$ case discussed below), the right-handed stop is  assumed to be very heavy (decoupled to be 2 TeV), which will result in a smaller mixing for a large range of $A_t \in [-4000, 4000]$ GeV. Furthermore, even if a large mixing in stop sector gives a much lighter  stop compared with the sbottom, this would potentially lead to a better signal in the stop sector.  The combination of the stop and sbottom signals, however,  does not depend on the mass difference between the stop and sbottom.  In the parameter space that we are considering with relatively small stop and sbottom mass difference, both channels contribute significantly to the combined reach.  In cases when the mass difference between the stop and sbottom is large, only one channel will contribute dominantly to the combined significance.

\begin{comment}
\begin{description}
\centering
\item[Significance: ] $\frac{S}{\sqrt{B+(10\%B)^2}}$,
\item[Confidence Level: ] $\frac{S}{\sqrt{S+B+(10\%B)^2}}$.
\end{description}
\end{comment}

\subsubsection{The Case of $\mu<0$: final states with a $Z$-boson}

For the case of $\mu<0$, the dominant decay channel of $\chi_2^0$ is $\nn\to Z \n$ instead \cite{Han:2013kza}. The leading signal under consideration for the pair production of sbottom and stop with $\tilde{b}_1 \to b \nn \to b Z \n$, $\tilde{b}_1^* \to t\chi_1^-\to bW^+\ W^-\n$ and $\tilde{t}_1\to t\nn\to bW^+Z\n$, $\tilde{t}_1^*\to b\chi_1^-\to bW^-\n$, is then 
 $$\tilde{b}_1\tilde{b}_1^*, \tilde{t}_1\tilde{t}_1^* \rightarrow bb\ WW\ Z\ \met \rightarrow \ell^+\ell^-\ bb\ jjjj\ \met.$$ 
The signal contains two $b$-jets, four light flavor jets, two same flavor, opposite sign leptons, and large missing energy.   The two leptons are used to reconstruct the $Z$ boson, which will significant reduce the SM backgrounds. 
%however, the cross section will also be suppressed due to the low branching ratio of the leptonical decay of the $Z$ boson. 
The dominant background is  $t \bar t$ plus one or two additional QCD jets.

We impose the basic event selection cuts as the previous case. We again optimize the cuts and divide the events into signal regions:
\begin{itemize}
\item $\slashed{E}_T$ to be above
100, 120, 140, 160 180, and 200 GeV.
\item $H_T$ to be above 400, 450, 500, 550, 600 GeV.
%\item $M_{\text{eff}}$, the effective mass of event, defined as scalar sum of transverse momentum $p_T$ of all visible final states in addition with Missing Energy:
%\begin{equation}
%M_{\text{eff}}=\sum_{\substack{all\ visible \\particles}}|p_T| + \slashed{E}_T,
%\end{equation}
%being 400, 450, 500, 550, 600, 650, 700, 750 and 800 \gev.
\item $M_{T2}$, the lepton-bashed transverse mass \cite{mt2_1,mt2_2,Cheng:2008hk}: 
\begin{equation}
\label{equ:MT2}
M_{T2}({\bf p}_T^{\ell_1},{\bf p}_T^{\ell_2},{\bf p}_T^{\rm miss})=\min_{{\bf p}_{T,1}^{\rm miss}+{\bf p}_{T,2}^{\rm miss}={\bf p}_T^{\rm miss}}\{\max\{M_T({\bf p}_T^{\ell_1},{\bf p}_{T,1}^{\rm miss}),M_T({\bf p}_T^{\ell_2},{\bf p}_{T,2}^{\rm miss})\}\}
\end{equation} 
to be above 75, 80, 85, 90 GeV.  
\item $\Delta M_{\ell\ell}=|M_{\ell\ell}-m_Z|$,  being less than 10 \gev.
%\item $M_{llb}$, the invariant mass of the lepton-pair and the hardest b-jet:
%\begin{equation}
%M_{llb}=\sqrt{(P_{l_1}+P_{l_2}+P_{b,1})^\mu(P_{l_1}+P_{l_2}+P_{b,1})_\mu}
%\end{equation}
%being 200, 280, 360, 440, 520, 600 and 680 \gev.
\item $N_j$ being at least 4, 5 and 6.
\item $N_b$ being at least 1 to suppress the enormous QCD backgrounds with light falvor jets.
\end{itemize}
 
\begin{figure}[tb] 
\includegraphics[width = 0.49 \textwidth]{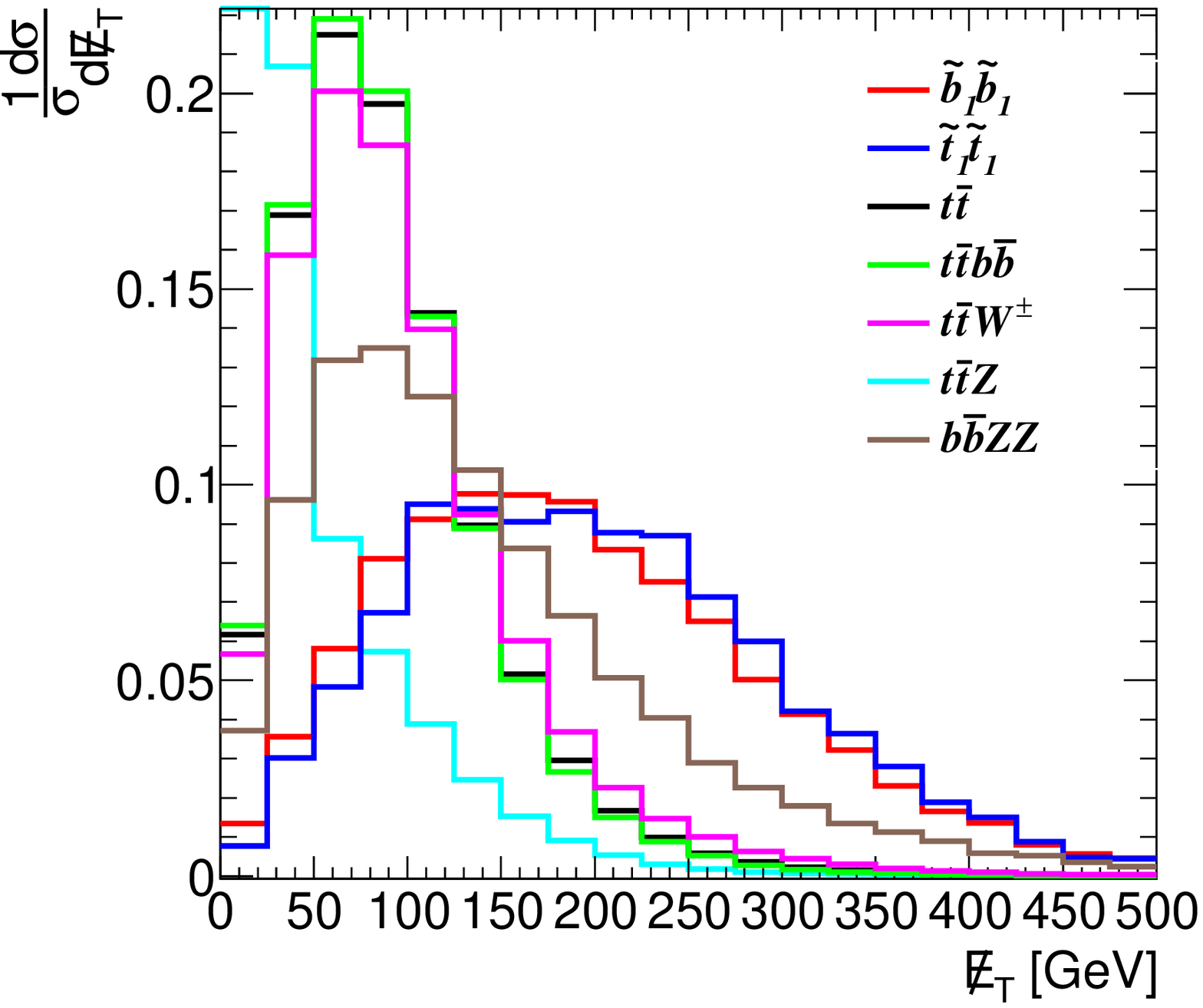}
\includegraphics[width = 0.49 \textwidth]{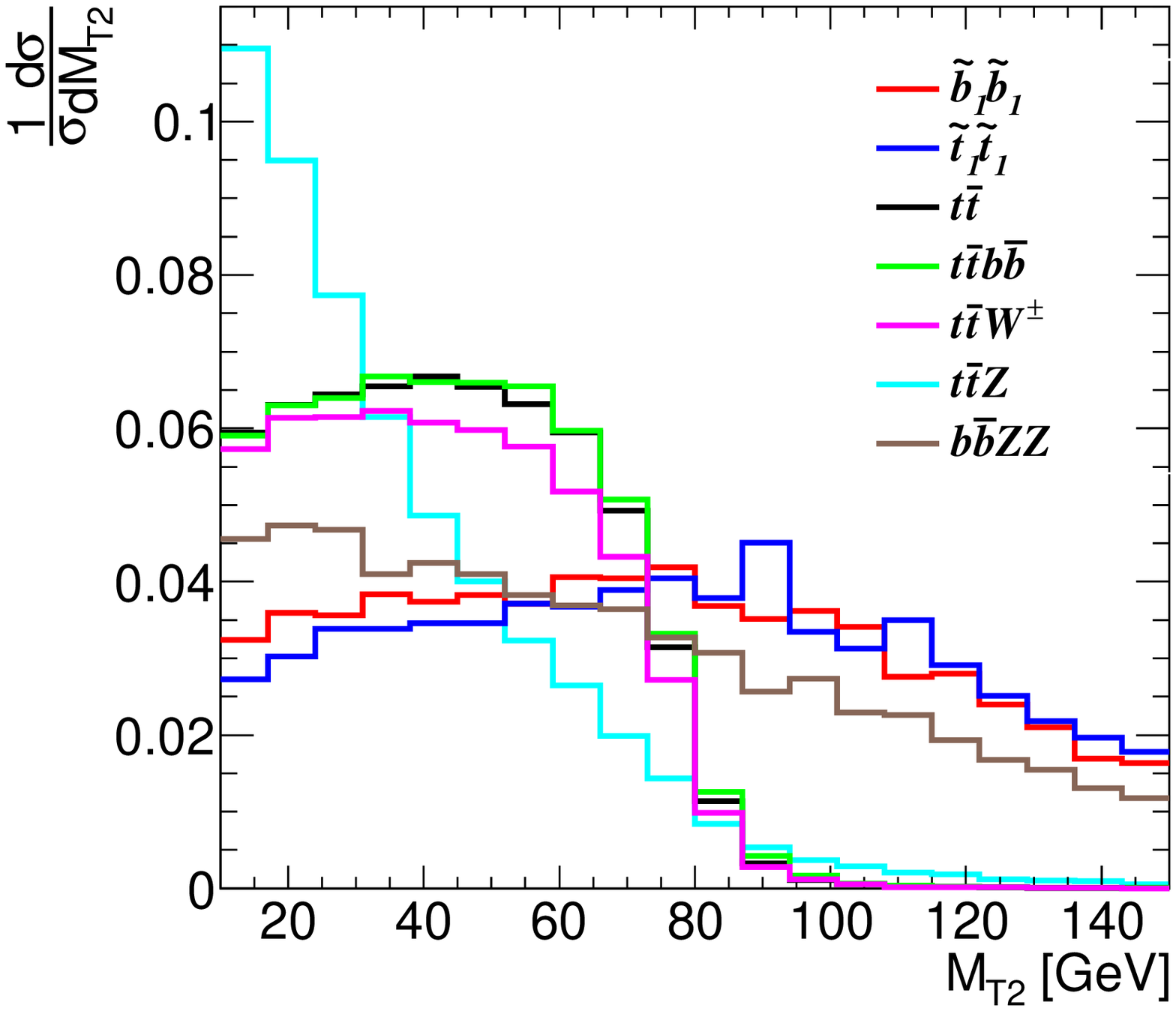}
\caption{Normalized distributions $\met$ (left panel) and $M_{T2}$ (right panel) for the signal $\tilde{b}_1\tilde{b}^*$ (red curve),  $\tilde{t}_1\tilde{t}_1^*$  (blue  curves) $\rightarrow bbWWZ\met \rightarrow \ell^+\ell^-\ bb\ jjjj\ \met$ after basic cuts with $m_{\tilde{b}_1}= 637$ GeV, $m_{\tilde{t}_1}= 634$ GeV, as well as SM backgrounds at the 14 TeV LHC.}
\label{fig:VariablesII}
\end{figure}

The normalized distributions of $\met$ and $M_{T2}$ for both the sbottom and stop signal, as well as the SM backgrounds are presented in Fig.~\ref{fig:VariablesII}.   The $\met$ distributions for the signal typically extend to larger values.  The $M_{T2}({\bf p}_T^{\ell_1},{\bf p}_T^{\ell_2},{\bf p}_T^{\rm miss})$ distributions for SM backgrounds with the lepton pair coming from leptonic $W$ decay are cut off at $m_W$, while the signal as well as $bbZZ$ background have much flatter $M_{T2}$ distributions.   Note that while the distribution of $bbZZ$ background is similar to that of the signal, the overall cross section for $bbZZ$ is negligibly small.
%too small to be of importance.  

%The backgrounds except $bbZZ$ are well separated from signal for $\slashed{E}_T$, however, due to the lack of jets and very low cross section for the process $bbZZ$, $bbZZ$ will be highly suppressed. And the peak value of $\slashed{E}_T$ for signal will be shifted to the right as the mass of LSP increases. The $M_{T2}({\bf p}_T^{l_1},{\bf p}_T^{l_2},{\bf p}_T^{\rm miss})$ is bounded sharply from above by the mass of the W boson for those background processes of which the lepton-pair is coming from the leptonic decay of two W boson. For other processes, including SUSY signal, the distribution of $M_{T2}$ is smooth over a larger range as indicated in the right panel of Figure.\ref{fig:VariablesII}.

%\Yongcheng{Although we don't use Mllb for cuts any more, this variable could indicate the mass of sbottom, shall we keep this variable for some discussion?}  For $M_{llb}$, in general, if we could identify the b jet which is the one produced together with $\chi_2^0$, hence with Z-boson, then the value of $M_{llb}$ should be a little less than the mass of sbottom (which is still larger than the masses of SM particles) due to the missing energy carried by $\chi_1^0$. We find that in this figure, $M_{llb}$ of signal is indeed larger than SM backgrounds in general. And if we can have more precision measurement about the momentum and energy of jets, the distribution of $M_{llb}$ will give us the information about the mass of $\tilde{b}_1$. 

Another interesting variable for the sbottom case is $M_{\ell\ell b}$, which is related to $m_{\tilde{b}_1}$ if the $b$ jet and the lepton pair from the same sbottom cascade decay chain $\tilde{b}_1 \to b \nn \to b Z \n$ can be identified.  While we will not use it for event selection in our analyses, $M_{\ell\ell b}$ distribution could provide information on $m_{\tilde{b}_1}$ as well as $m_{\nn}$ if a sbottom signal is discovered.

The advanced cuts and the corresponding cumulative cut efficiencies as well as the cross sections for sbottom and stop signal for BP2 with  $\mu < 0$ and SM backgrounds before and after cuts are  given in Table \ref{tab:CUTSResultII}.
The dominant SM background is $t\bar{t}$ plus jets.     A significance of   about 12$\sigma$ (8.7$\sigma$) can be reached for $\tilde b_1$ \ ($\tilde t_1$) for the benchmark point at the 14 TeV LHC with 300 ${\rm fb}^{-1}$ luminosity,  including 10\% systematic error. 

\begin{comment} The experimental search significance for this case is much smaller than the case of $\mu>0$ due to the small branching ratio of Z-boson decaying into two leptons which leads to the smallness of cross section of signal process for the case of $\mu<0$, however, it can still reach $8.71\sigma$ after all the cuts.
\end{comment} 

 \begin{table}[tb]
\resizebox{\textwidth}{!}{
\begin{tabular}{|c|c|c|c|c|c|c|c|c|c|}
\toprule[1pt]
Process & $\sigma$ (fb) & Basic &$\slashed{E}_T>$ & $H_{T}>$ & $M_{T2}>$ & $\Delta M_{ll}<$ & $N_j\geq$ & $N_b\geq$ & $\sigma$ (fb) \\
  &  & cuts & $175\text{ GeV}$ &$400\text{ GeV}$ &$90\text{ GeV}$ &$10\text{ GeV}$ &$4$ &$1$ &after cuts\\
\toprule[1pt]
%$\tilde{b}_1 \tilde{b}_1$&6.5&28\% & 11\% & 8.4\% & 3.2\% & 2.9\% & 2.9\% & 2.2\% & $1.4\times10^{-1}$ \\
$\tilde{b}_1 \tilde{b}_1$&2.1&32\% & 17\% & 16\% & 5.8\% & 5.3\% & 5.3\% & 4.2\% & $8.8\times10^{-2}$ \\

\hline
%$\tilde{t}_1 \tilde{t}_1$ & 4.02 & 21\% & 8.6\% & 5.8\% & 3.0\% & 2.8\% & 2.8\% & 2.0\% & $8.1\times10^{-2}$ \\ 
$\tilde{t}_1 \tilde{t}_1$ & 1.8 & 27\% & 16\% & 11.2\% & 4.9\% & 4.4\% & 4.4\% & 3.6\% & $6.5\times10^{-2}$ \\
\hline
$t\bar{t}$&33,000&1.3\% & 0.09\% & 0.06\% & $5.0\times10^{-6}$ & $4.9\times10^{-7}$ & $4.9\times10^{-7}$ & $3.7\times10^{-7}$ & $1.2\times10^{-2}$ \\
\hline
$t\bar{t}Z$&71&11\% & 0.25\% & 0.16\% & $5.8\times10^{-5}$ & $4.2\times10^{-5}$ & $4.2\times10^{-5}$ & $2.7\times10^{-5}$ & $1.9\times10^{-3}$ \\
\hline
$t\bar{t}b\bar{b}$&400&3.2\% & 0.20\% & 0.12\% & $1.4\times10^{-5}$ & $2.0\times10^{-6}$ & $2.0\times10^{-6}$ & $1.8\times10^{-6}$ & $6.9\times10^{-4}$ \\
\hline
$t\bar{t}ZZ$&0.16&16\% & 0.86\% & 0.64\% & 0.31\% & 0.27\% & 0.27\% & 0.18\% & $3.0\times10^{-4}$ \\
\hline
%$t\bar{t}W^{\pm}$&47&3.2\% & 0.34\% & 0.19\% & $1.2\times10^{-5}$ & $1.3\times10^{-6}$ & $1.3\times10^{-6}$ & $1.0\times10^{-6}$ & $4.7\times10^{-5}$ \\
%\hline
%$b\bar{b}W^{\pm}W^{\mp}$&62&0.34\% & 0.05\% & $3.0\times10^{-4}$ & $3.8\times10^{-6}$ & $2.5\times10^{-7}$ & $2.5\times10^{-7}$ & $2.5\times10^{-7}$ & $1.6\times10^{-5}$ \\
%\hline
$b\bar{b}ZZ$&2.3&0.39\% & 0.11\% & 0.06\% & $2.9\times10^{-4}$ & $2.6\times10^{-4}$ & $2.6\times10^{-4}$ & $2.1\times10^{-4}$ & $4.8\times10^{-4}$ \\
\hline
&&\multicolumn{6}{r}{$\sqrt{s}=14\text{ TeV}\qquad \int L \ dt = 300 \ {\rm fb}^{-1}\qquad $} &  &  \\
&&\multicolumn{6}{r}{${S\over \sqrt{B+(10\%B)^2}}$ = 12 \ (8.7) \quad for \quad $\tilde b_1$ \ ($\tilde t_1$)} &  & \\
%&&\multicolumn{6}{r}{ } & sbottom & 11.8 \\
\bottomrule[1pt]
\end{tabular}
}
\caption{ Cut efficiencies and cross sections before and after cuts for the signal $\tilde{b}_1\tilde{b}^*_1, \tilde{t}_1\tilde{t}_1^* \rightarrow bbWWZ\met \rightarrow \ell^+\ell^-\ bb\ jjjj\ \met$,
for BP2  in Table \ref{tab:massParameters} for $\mu <0$, as well as dominant SM backgrounds at the 14 TeV LHC.  The significance is  obtained  for  $\int L \ dt$ = 300 ${\rm fb}^{-1}$ with 10\% systematic error combining both sbottom and stop signals. }  
\label{tab:CUTSResultII}
\end{table}

\begin{figure}[tb]
\minigraph{7cm}{-0.3in}{(a)}{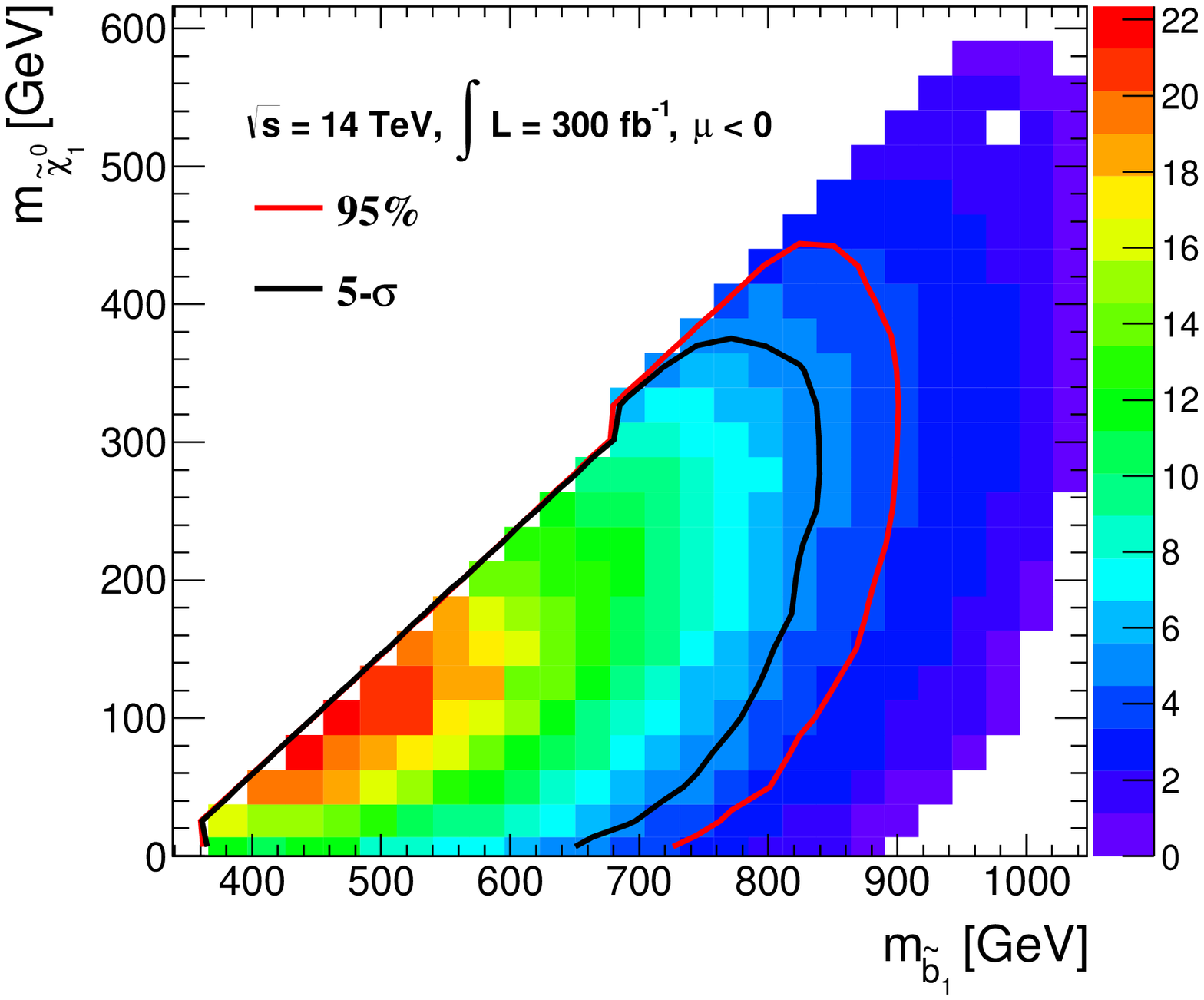}
\minigraph{7cm}{-0.3in}{(b)}{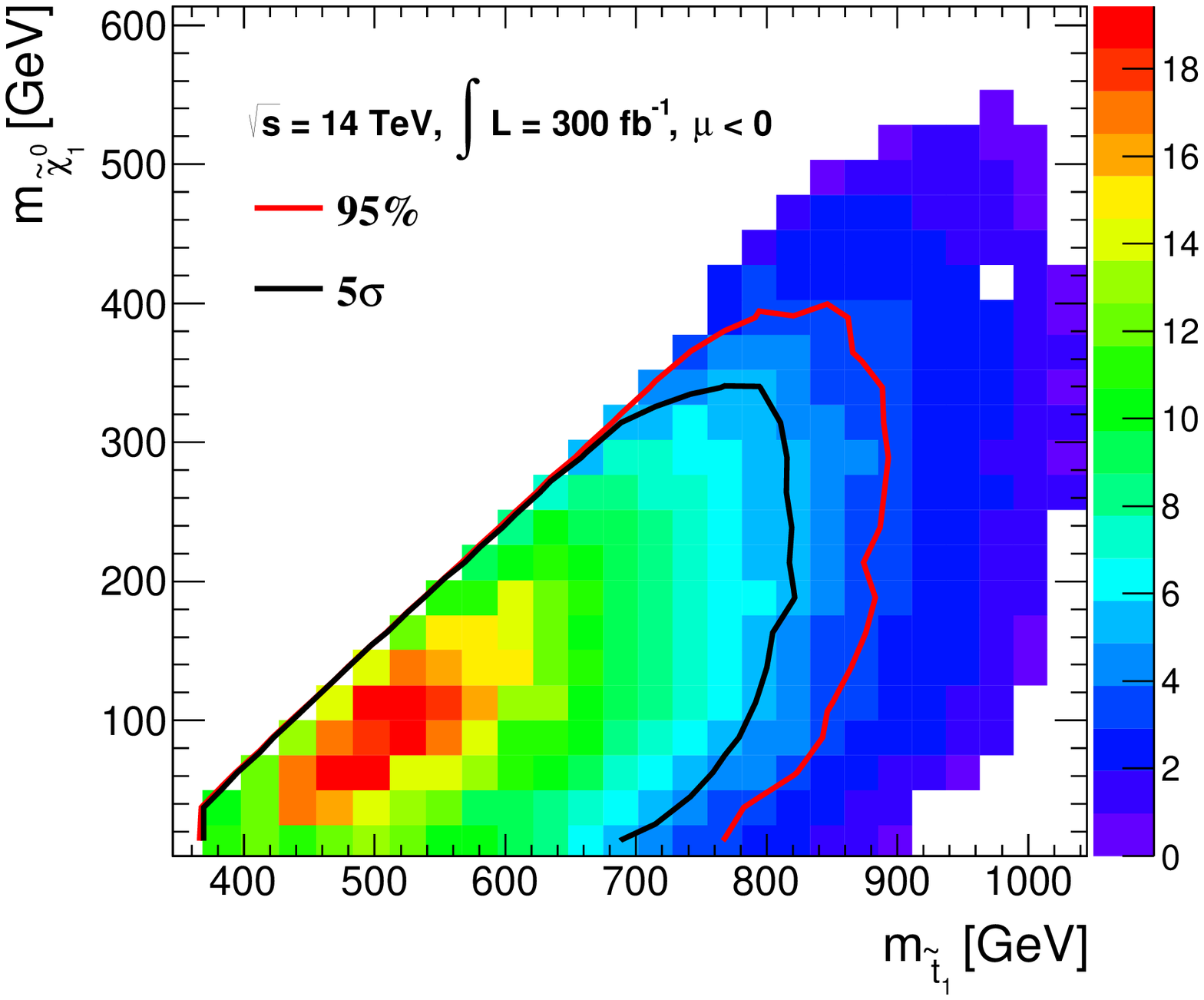}
\minigraph{7cm}{-0.3in}{(c)}{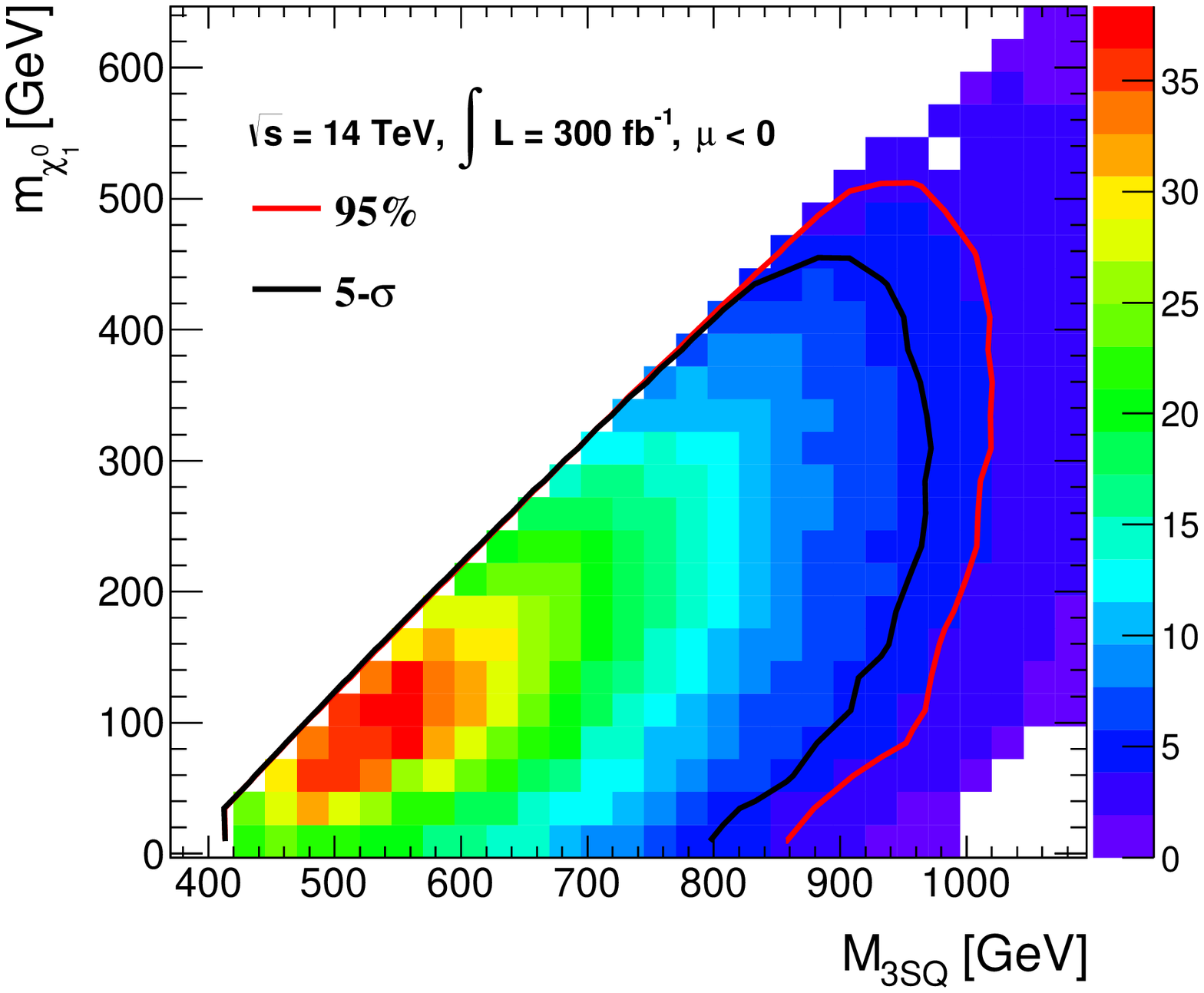}
\caption{
Signal significance contours for $\tilde{b}_1\tilde{b}^*, \tilde{t}_1\tilde{t}_1^* \rightarrow bbWWZ\met \rightarrow \ell^+\ell^- bbjjjj\met$ final states for 14 TeV LHC with ${\int L \ dt}\ = \ 300\text{ fb}^{-1}$ luminosity. The 5$\sigma$ discovery reach (black curves) and 95\% C.L. exclusion limit (red curves) are show in the (a) $m_{\tilde{b}_1}-m_{\chi_1^0}$ plane, in the (b) $m_{\tilde{t}_1}-m_{\chi_1^0}$ plane,  and the combined reach in the (c) $M_{3SQ}-m_{\chi_1^0}$ plane.}
%  for $\mu<0$, at  14 TeV LHC with    300 ${\rm fb}^{-1}$ integrated luminosity. } % \Shufang{ Can sbottom reach gets better if Mllb is used in addition?} \Huanian{For both cases, sbottom reach will be better than stop reach.}  \Shufang{The plot shows stop reach is better than that of sbottom.} \Yongcheng{The 5-sigma reach could get a little better when adding Mllb, but the 5-sigma reach line would roughly cover with 95\% reach line. } \Shufang{YC, I don't understand your comment.  How can 5 sigma reach over 95\% exclusion line? or you are talking about two different sets of cut here?}\Yongcheng{We use different function to calculate the 5-sigma reach and 95\% exclusion. $5-\sigma=\frac{S}{\sqrt{B+(10\%B)^2}}$ and $95\%=\frac{S}{\sqrt{S+B+(10\%B)^2}}$. And  when we add Mllb further, the events left for both signal and backgound are so small that the 5-sigma line can cover the 95\% line in some region.}}
\label{fig:SGCLII}
\end{figure}

Signal significance contours are shown in Fig.~\ref{fig:SGCLII} with the 5$\sigma$ discovery reach (black curve) and 95\% C.L. exclusion limit (red curve) for 14 TeV LHC with 300 ${\rm fb}^{-1}$ integrated luminosity, in the (a) $m_{\tilde{b}_1}-m_{\chi_1^0}$ plane, 
(b) $m_{\tilde{t}_1}-m_{\chi_1^0}$ plane, and  (c) $M_{3SQ}-m_{\chi_1^0}$ plane.   For massless $\n$, sbottom (stop) masses up to 650 (680) $\gev$ can be discovered and 720 (760) will be excluded at 95\% C.L. if there is no signal over SM backgrounds being found.   For moderate mass of $\n$ around 200 $\sim$ 300 GeV, the 5$\sigma$ dicovery can reach up to 820 (840) GeV, and the 95\% exclusion limit can go up to 890 (900) GeV for sbottom (stop). The combined reach of the stop and sbottom is shown in  Fig.~\ref{fig:SGCLII} (c)  in $M_{3SQ}$ versus $m_{\chi_1^0}$ plane.    About  980 GeV can be achieved in $M_{3SQ}$ for the 5$\sigma$ discovery reach and about 1025 GeV for the 95\% C.L. exclusion.
The experimental reach for the case of $\mu<0$ is lower than that for the case of $\mu>0$.

%Event samples are generated  using Madgraph MG5$\_$aMC$\_$V2$\_$2$\_$1 \cite{Alwall:2014hca},   processed   through Pythia 6.420 \cite{oai:arXiv.org:hep-ph/0603175} for   fragmentation and hadronization and then through Delphes-3.1.2 \cite{deFavereau:2013fsa} with the Snowmass combined LHC detector card  \cite{Anderson:2013kxz} for  detector simulation.  Both the SM backgrounds and the stop pair production signal are normalized to theoretical cross sections, calculated including higher-order QCD corrections \cite{oai:arXiv.org:1006.4771, Broggio:2013uba, oai:arXiv.org:0804.2800, oai:arXiv.org:0905.0110, oai:arXiv.org:hep-ph/0211352, oai:arXiv.org:1204.5678, oai:arXiv.org:0804.2220}.      For the signal process, we scan the parameter range of $M_{3SQ}=400 - 1100$ GeV with step size of 25 GeV, and $M_1= 3 - 750$ GeV with step size of  25 GeV.  We fix $M_2$ with $M_2=M_1+$ 150 GeV.
 
%\Shufang{Huanian, add your part of stop as well as combined results.}

%%%%%%%%%%%%%%%%%%%%%%%%%%%%%%
\subsection{Signature of $\tilde{b}_1\sim\tilde{b}_R$}

To complete our exploration for the sbottom signal, we consider another scenario with the low energy mass spectrum containing a light mostly right-handed sbottom, a Bino-like LSP and Higgsino-like NLSPs. Here, the sign of $\mu$ does not affect the decay modes of sbottom and neutralinos. The typical benchmark point is listed in Table.~\ref{tab:RHsbottomBM}, the corresponding branching fractions are listed in Table.~\ref{tab:DecayBRBP3}. Other soft SUSY breaking parameters are decoupled by setting them to be at 2 TeV. In this scenario, the right-handed sbottom couples to the Bino and Higgsino through the ${\rm U(1)}_Y$ or the bottom Yukawa couplings, which results in the sbottom dominantly decaying to $b \chi_1^0$ due to the large phase space, followed by the channel $t \chi_1^\pm$ when it is kinematically open. We will focus on the signal reach of the sbottom pair production 
$$\tilde b_1 \tilde b_1^* \to b \chi_1^0\ t \chi_1^\pm \to b \chi_1^0\ t W^\pm \chi_1^0 \to \ell\ bb\ jj\ \met.$$ 
The SM backgrounds are somewhat similar to that of the $\mu>0$ case of left-handed sbottom with fewer jets. We also include vector bosons plus additional jets as another background \cite{Avetisyan:2013onh}.
%, since there is less b-jets in the final states. 

\begin{table}[tb]
\begin{tabular}{|c|c|c|c|c|c|c||c|c|c|c|c|c|}
\hline
&$M_1$ & $M_2$ & $M_{3SD}$& $A_t$ & $\mu$ & $\tan\beta$ & $m_{\chi_1^0}$  & $m_{\chi_1^\pm}$ &$m_{\tilde{b}_1}$ & $m_h$ \\
\hline
%150 & 300 & 550 & 2950 & +2000 & 10 &151 & 319 & 319 & 526 & 538 & 125 \\
BP3&150 & 2000 & 650 & 2895 & 300 & 10 & 145 & 307 & 635 & 125 \\
\hline
%150 & 300 & 550 & 2950 & $-$1300 & 10 & 152 & 322 & 321 & 526 & 525 & 124 \\
%150 & 300 & 650 & 2950 & $-$1300 & 10 & 152 & 322 & 322 & 637 & 634 & 125 \\
%\hline
\end{tabular}
\caption{MSSM parameters and mass spectrum of SUSY particles for the the benchmark point in the case of Right-handed sbottom.   All masses are in units of GeV. }
\label{tab:RHsbottomBM}
\end{table}

\begin{table}[tb]
\begin{tabular}{|c|c|c||c|c|}
\toprule[1pt]
&Decay Channel & BR & Decay Channel & BR  \\
\midrule[1pt]
\multirow{2}{*}{BP3}&$\tilde{b}_1\to b\n$ & 58\% & $\chi_1^\pm\to \chi_1^0 W^\pm$ & 100\% \\
\cline{2-5}
&$\tilde{b}_1\to t\chi_1^-$ & 18\% & & \\
\bottomrule[1pt]
\end{tabular}
\caption{Decay channels and the corresponding branching fractions of $\tilde{b}_1$ and $\chi_1^\pm$ for the benchmark point, which corresponds to the case of Right-handed sbottom. } % \Shufang{ Do we need the table? $\mu>0$ and $\mu<0$ are very similar for stop and sbottom decay, with the only difference being chi20 decay.}}
\label{tab:DecayBRBP3}
\end{table}

We scan over a broad mass parameter space: $M_1$ from 3 GeV to 800 GeV in step of 30 GeV, $M_{3SD}$ from 400 GeV to 1180 GeV in step of 30 GeV, $\mu$ is fixed to be $\mu=M_1+150\text{ GeV}$. We further require that $m_{\tilde{b}_1}>m_{\chi_1^\pm}+m_t$ so that the decay channel $\tilde{b}_1\to t\chi_1^\pm$ is kinematically accessible. Since the final state particles are more stiff than the previous cases with cascade decays, 
we apply stronger basic cuts than before. The basic event selection cuts are

\begin{itemize}
\item Jet:
\begin{equation}
\label{equ:BasicCutsJets2}
|\eta_{j}|<2.5 ,\quad p_{T}^{j} > 40 \text{ GeV} ,\quad \Delta\phi_{j,\met} > 0.8.
\end{equation}
%where $\Delta\phi_{j,\slashed{E}_T}$ is azimuthal angle between the jet and missing transverse energy.
\item  Lepton:
\begin{equation}
\label{equ:BasicCutsLep2}
|\eta_{\ell}|<2.5 ,\quad p_{T}^{\ell} > 30 \text{ GeV} ,\quad \Delta R_{\ell j} > 0.4.
\end{equation}
%where the $\Delta R_{\ell j}$ is the distance in the $\phi$-$\eta$ plane: $\Delta R = \sqrt{\Delta\phi^2+\Delta\eta^2}$,  between the lepton and the jet satisfying Eq.~(\ref{equ:BasicCutsJets2}).
\item at least three jets satisfying requirement Eq.~(\ref{equ:BasicCutsJets2}), within which at least one $b$-tagged, and exactly one lepton satisfying requirement Eq.~(\ref{equ:BasicCutsLep2}).
\item the leading $b$-jet $p_T$ is required to be larger than 100 GeV since one $b$-jet originates directly from a heavy sbottom decay.
\end{itemize}

Besides the basic event selection cuts, we apply the same advanced event selection cuts in the signal regions ($H_T$, $\met$, $M_T$, $N_j$ and $N_b$) as in Sec.~\ref{sec:LHsbottom}, and optimize them for different mass parameters. In Table \ref{tab:RH-CutEff}, we list the cross section before and after above cuts and also the efficiency after every cut for the benchmark point listed in Table \ref{tab:RHsbottomBM}.

\begin{table}[tb]
\resizebox{\textwidth}{!}{
\begin{tabular}{|c|c|c|c|c|c|c|c|c|}
\toprule[1pt]
Process & $\sigma$ (fb) & Basic &$\slashed{E}_T>$ & $H_{T}>$ & $M_{T}>$ & $N_j\geq$ & $N_b\geq$ & $\sigma$ (fb) \\
  &  & Cuts & $200\text{ GeV}$ &$500\text{ GeV}$ &$160\text{ GeV}$ &$4$ &$1$ &after Cuts\\
\toprule[1pt]
$\tilde{b}_1\tilde{b}_1$&9.7&30\% & 20\% & 14\% & 8.1\% & 5.6\% & 5.6\% & $5.4\times10^{-1}$ \\
\hline
$t\bar{t}$& 260,000&5.3\% & 0.14\% & $4.7\times10^{-4}$ & $1.6\times10^{-6}$ & $8.1\times10^{-7}$ & $8.1\times10^{-7}$ & $2.1\times10^{-1}$ \\
\hline
$t\bar{t}b\bar{b}$&2,300&13\% & 0.4\% & 0.2\% & $3.7\times10^{-5}$ & $2.6\times10^{-5}$ & $2.6\times10^{-5}$ & $6.1\times10^{-2}$ \\
\hline
$t\bar{t}h$&100&20\% & 1\% & 0.7\% & $7.8\times10^{-5}$ & $5.2\times10^{-5}$ & $5.2\times10^{-5}$ & $5.3\times10^{-3}$ \\
\hline
$t\bar{t}Z$&230&14\% & 0.7\% & 0.5\% & $8.1\times10^{-5}$ & $4.5\times10^{-5}$ & $4.5\times10^{-5}$ & $1\times10^{-2}$ \\
\hline
$t\bar{t}W^{\pm}$&224&11\% & 0.7\% & 0.5\% & $6.6\times10^{-5}$ & $3.4\times10^{-5}$ & $3.4\times10^{-5}$ & $7.6\times10^{-3}$ \\
\hline
$Vjj$ &$3.7\times10^{7}$ & $4.8\times10^{-5}$ & $2.9\times10^{-6}$ & $1.8\times10^{-6}$ & $2.9\times10^{-9}$ & $1\times10^{-9}$ & $1\times10^{-9}$ & $3.8\times10^{-2}$ \\
\hline
&&\multicolumn{5}{r}{$\sqrt{s}=14\text{ TeV}\qquad L=300{\rm fb}^{-1}\qquad S/\sqrt{B+(10\%B)^2}$} &  & 11.4 \\
\bottomrule[1pt]
\end{tabular}
}
\caption{Cut efficiencies and cross sections before and after cuts for the signal $\tilde{b}_1\tilde{b}^*_1 \rightarrow bbWW\met \rightarrow \ell\ bb\ jj\ \met$, for BP3 in Table \ref{tab:RHsbottomBM}, 
as well as SM backgrounds at the 14 TeV LHC.  The significance is  obtained  for  $\int L dt$ = 300 ${\rm fb}^{-1}$ with 10\% systematic error. }
\label{tab:RH-CutEff}
\end{table}

\begin{figure}[tb]
\includegraphics[scale=0.40]{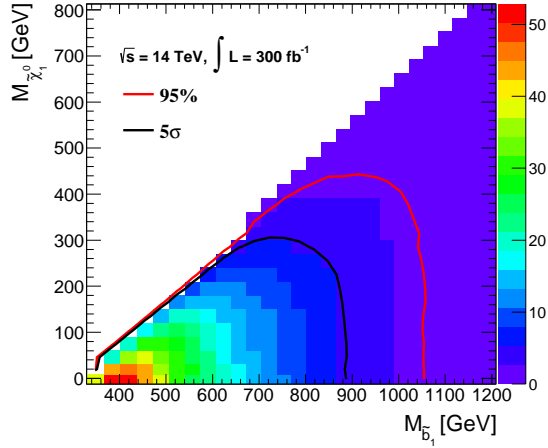}
\caption{
Signal significance contours for 
$\tilde{b}_1\tilde{b}_1^* \rightarrow bbWW\met \rightarrow \ell^\pm\ bb\ jj\ \met$ final state for the right-handed sbottom in the $m_{\tilde{b}_1}-m_{\chi_1^0}$ plane for 14 TeV LHC with ${\int L \ dt}\ = \ 300\text{ fb}^{-1}$ luminosity.
The 5$\sigma$ discovery reach (black curves) and 95\% C.L. exclusion limit (red curves) are shown.
 }
\label{fig:rhsb-reach}
\end{figure}

Signal significance contours are shown in Fig.~\ref{fig:rhsb-reach} with the 5$\sigma$ discovery reach (black curve) and 95\% C.L. exclusion limit (red curve) for 14 TeV LHC with 300 ${\rm fb}^{-1}$ integrated luminosity, in the $m_{\tilde{b}_1}-m_{\chi_1^0}$ plane.
%
%Fig.~\ref{fig:rhsb-reach} illustrates the 95\% C.L. exclusion as well as 5$\sigma$ reach for right-handed sbottom in $bbW(\ell^\pm\nu)W(jj)+\met$ channel at $\sqrt{s}=14$ TeV and 300 ${\rm fb}^{-1}$ integrated luminosity. 
%
For a large range of mass of $\n$ (from massless to about 300 GeV), sbottom masses up to about 880 GeV can be discovered and 1050 GeV will be excluded at 95\% C.L. if there is no further signal over SM backgrounds being found. The reach at lower mass of $\n$ is better than that of left-handed case, since lowering the mass of $\n$ will increase the $p_T$ of the $b$-jet produced together with $\n$, and this effect is suppressed in the left-handed case where the leading $b$-jets is produced together with $\nn$ or $\cha$ which are always heavier than $\n$.

\section{Summary and Conclusion}
\label{sec:conclusion}

In this paper, we stress the point that in a realistic situation in a generic MSSM, the sbottom decay can be far from $100\%$ to a specific channel, as assumed in most of the current studies and all the LHC sbottom searches,  which is only true for the Bino-LSP and either  the left-handed sbottom (or a nearly degenerate stop) being the NLSP, or the right-handed sbottom (or Wino) being the NLSP.
On a more general ground, sbottom decays lead to a much richer pattern. The inclusion of the other decay channels will significantly weaken the current sbottom search limits and in the mean time open new decay modes for alternative discovery channels for sbottom searches. 

We studied in detail the sbottom decay patterns in a few representative SUSY mass scenarios.  For the left-handed sbottom, we found that 
\begin{itemize}
\item[(1)] in the Wino-NLSP case, see Fig.~\ref{fig:BRofsb1}(a),  ${\rm BR}(\tilde{b}_1\rightarrow b\nn) \sim {\rm BR}(\tilde{b}_1\rightarrow t\chi_1^\pm) \sim 50\%$ while ${\rm BR}(\tilde{b}_1\rightarrow b \n) \sim 2\%$. 
\item[(2)] in the Higgsino-NLSP case, see Fig.~\ref{fig:BRofsb1}(b), 
  $\tilde{b}_1\rightarrow t\chi_1^\pm$ dominates while $\tilde{b}_1\rightarrow b\chi_{1,2,3}^0$ are all suppressed. 
\item[(3)] in mixed NLSP cases, see Figs.~\ref{fig:BRofsb1}(c) and 1(d), 
${\rm BR}(\tilde{b}_1\rightarrow b\nn) \sim {\rm BR}(\tilde{b}_1\rightarrow t\chi_1^\pm) \sim 
{\rm BR}(\tilde{b}_1\rightarrow t\chi_2^\pm) \sim 30\%$ and ${\rm BR}(\tilde{b}_1\rightarrow b \n) \sim 3\%$ when $|\mu| > M_2$; 
while ${\rm BR}(\tilde{b}_1\rightarrow t\chi_1^\pm) \sim 
{\rm BR}(\tilde{b}_1\rightarrow t\chi_2^\pm) \sim 30\%$ and 
${\rm BR}(\tilde{b}_1\rightarrow b\chi_1^0) < 10\%$ 
when $M_2> |\mu|$.
\end{itemize}
 
For the right-handed sbottom, see Fig.~\ref{fig:BRofR}, decays of $\tilde{b}_1\rightarrow b\chi_1^0$ dominates for  the case of Bino-LSP with Wino-NLSP.  In the case of Bino-LSP with Higgsino-NLSP, however, the branching fraction of  $\tilde{b}_1\rightarrow b\chi_1^0$ is reduced to about 40\%$-$60\%, while  $\tilde{b}_1\rightarrow t\chi_1^\pm$ is about 20$-$30\%, followed by  $\tilde{b}_1\rightarrow b\chi_{2,3}^0$ of about 10\% each. 
 
We analyzed in detail the sbottom pair production signals with the mixed decay channels.
% \Tao{for the above cases (1) and (2), which serve as representatives for case (3).} 
We focus on the search sensitivity at the 14 TeV LHC with a 300 $\rm fb^{-1}$ integrated luminosity. We scanned over a large SUSY mass parameter region and performed semi-realisc detector simulations. For the left-handed sbottom $\bL$ pair production,  we focused on the scenario of Bino-LSP with Wino-NLSP.  With  one sbottom decaying via $\tilde{b}\rightarrow b \chi_2^0$ and the other sbottom decaying via $\tilde{b} \rightarrow t \chi_1^\pm$, we found that
\begin{itemize}
\item[$\bullet$] 
With $\chi_2^0 \rightarrow h \chi_2^0\ (\mu >0)$ and $\chi_1^\pm \rightarrow W^\pm \chi_1^0$, the leading signal is the $bbbb\ jj\ \ell+\met$ final state.  From Fig.~\ref{fig:SGCLI}(a), we see that 
a 5$\sigma$ discovery can be made up to 920 GeV, and 
the 95\% C.L exclusion limit can reach up to 1050 GeV for this Higgs channel. 
The reach of the combined sbottom and stop signals of the same final states 
%$bbbb\ jj\ \ell+\met$ 
is about 120 GeV higher, as shown in Fig.~\ref{fig:SGCLI}(b).
\item[$\bullet$] 
With $\chi_2^0 \rightarrow Z \chi_2^0\ (\mu<0)$ and $\chi_1^\pm \rightarrow W^\pm \chi_1^0$, 
we studied the reach of $bb\ jjjj\ \ell\ell+\met$ final state. As seen from Fig.~\ref{fig:SGCLII},
a 5$\sigma$ discovery can be made up to 840 GeV, and 
the 95\% C.L exclusion limit can reach up to 900 GeV for the $Z$ channel. 
The 5$\sigma$ discovery potential of the combined sbottom and stop signals can reach up to 980 GeV, and the 95\% exclusion limit is about 1025 GeV.  
\end{itemize}
%\item[$\bullet$] 
We also studied the signal for the right-handed sbottom $\bR$  in the scenario of Bino-LSP with Higgsino-LSP. With one sbottom decaying via $\tilde{b}\rightarrow b \chi_1^0$, and the other sbottom decaying via $\tilde{b} \rightarrow t \chi_1^\pm$, we found that the reach of $bbjj\ell+\met$ 
%and $bbjj\ell\ell+\met$ 
final states can lead to
%be stronger than those for $\bL$. 
a 5$\sigma$ discovery up to 900 GeV, and 
the 95\% C.L exclusion limit up to 1060 GeV, 
as shown in Fig.~\ref{fig:rhsb-reach}. 
 Including the other commonly studied channels, $b\chi_1^0 \bar b \chi_1^0,\ b\chi_2^0 \bar b \chi_2^0$ and $t \chi_1^- \bar t \chi_1^+$ would help increase the overall search sensitivity, but we did not repeat the analyses as listed in Table I.

 \acknowledgments
 %We would like to thank xxx  for helpful discussions.   
%
The work of T.H.~was supported in part by the US Department of Energy under Grant~DE-FG02-95ER-40896 and in part by PITT PACC.
The work of S.S.~and H.Z.~was supported by DoE under Grant~DE-FG02-04ER-41298. 

\bibliography{bibliography}
 
\end{document}